\documentclass[longauth]{aa}  

\usepackage{graphicx}
\usepackage{txfonts}
\usepackage{hyperref}
\usepackage[normalem]{ulem}

\begin{document} 

\title{The all-sky PLATO input catalogue\thanks{
The catalogue described in this article is only available in electronic form at MAST as a High Level Science Product via
\url{https://dx.doi.org/10.17909/t9-8msm-xh08}
and
\url{https://archive.stsci.edu/hlsp/aspic}, in the
SSDC tools page (\url{https://tools.ssdc.asi.it/asPICtool/}) and at the CDS via anonymous ftp to
cdsarc.u-strasbg.fr (130.79.128.5) or via \url{http://cdsarc.u-strasbg.fr/cgi-bin/viz-bin/cat/J/A+A/vol/page}
}
}
   

\author{
M.~Montalto\inst{1,2}\fnmsep\thanks{E-mail: marco.montalto@unipd.it},
G.~Piotto\inst{1,2},
P.~M.~Marrese\inst{3,4,},
V.~Nascimbeni\inst{1,2},
L.~Prisinzano\inst{5},
V.~Granata\inst{1,2},
S.~Marinoni\inst{3,4},
S.~Desidera\inst{2},
S.~Ortolani\inst{1,2},
C.~Aerts\inst{14,15,16},
E.~Alei\inst{6},
G.~Altavilla\inst{3,4},
S.~Benatti\inst{5},
A.~B\"orner\inst{7},
J.~Cabrera\inst{8},
R.~Claudi\inst{2},
M.~Deleuil\inst{12},
M.~Fabrizio\inst{3,4},
L.~Gizon\inst{17,18,19},
M.~J.~Goupil\inst{9},
A.~M.~Heras\inst{10},
D.~Magrin\inst{2},
L.~Malavolta\inst{1,2},
J.~M.~Mas-Hesse\inst{13},
I.~Pagano\inst{11},
C.~Paproth\inst{7},
M.~Pertenais\inst{7},
D.~Pollacco\inst{20,21},
R.~Ragazzoni\inst{1,2},
G.~Ramsay\inst{24},
H.~Rauer\inst{8,22},
S.~Udry\inst{23}
}

\institute{
Dipartimento di Fisica e Astronomia "Galileo Galilei", Universit\'a di Padova, Vicolo dell'Osservatorio 3, Padova IT-35122, Italy
\and
Istituto Nazionale di Astrofisica - Osservatorio Astronomico di Padova, Vicolo dell'Osservatorio 5, 35122, Padova, Italy
\and
Istituto Nazionale di Astrofisica - Osservatorio Astronomico di Roma, Via Frascati 33, 00078 Monte Porzio Catone, Roma, Italy
\and
Space Science Data Center - ASI, Via del Politecnico snc, 00133 Roma, Italy
\and
Istituto Nazionale di Astrofisica – Osservatorio Astronomico di Palermo, P.zza del Parlamento 1, 90134 Palermo, Italy
\and
ETH   Z\"urich,   Institute   for   Particle   Physics   and Astrophysics,Wolfgang-Pauli-Str. 27, 8093 Z\"urich, Switzerland
\and
Deutsches  Zentrum  f\"ur  Luft-  und  Raumfahrt  (DLR),  Institut  f\"ur  Optische  Sensorsysteme,  Rutherfordstra{\ss}e  2,  12489  Berlin-Adlershof, Germany
\and
Deutsches Zentrum f\"ur Luft- und Raumfahrt (DLR), Institut f\"ur Planetenforschung, Rutherfordstra{\ss}e 2, 12489 Berlin-Adlershof, Germany
\and
LESIA, CNRS UMR 8109, Universit\'e Pierre et Marie Curie, Universit\'e Denis Diderot, Observatoire de Paris, F-92195 Meudon, France
\and
European Space Agency (ESA), European Space Research and Technology Centre (ESTEC), Keplerlaan 1, 2201 AZ Noordwijk, The Netherlands
\and
Istituto Nazionale di Astrofisica - Osservatorio Astrofisico di Catania, Via S. Sofia 78,I-95123, Catania, Italy
\and
Aix-Marseille   Universit\'e,   CNRS,   CNES,   Laboratoire   d’Astrophysique   de   Marseille,Technop\^{o}le  de  Marseille-Etoile,  38,  rue  Fr\'ed\'eeric  Joliot-Curie  F-13388  Marseille  cedex  13,France
\and
Centro de Astrobiolog\'{\i}a (CSIC-INTA), Depto. de Astrof\'{\i}sica, Madrid, Spain
\and
Institute of Astronomy, KU Leuven, Celestijnenlaan 200D, 3001, Leuven, Belgium
\and
Department of Astrophysics, IMAPP, Radboud University Nijmegen, 6500 GL, Nijmegen, The Netherlands 
\and
Max Planck Institute for Astronomy, Koenigstuhl 17, 69117, Heidelberg, Germany
\and
Max-Planck-Institut f\"ur Sonnensystemforschung, Justus-von-Liebig-Weg~3, 37077~G\"ottingen, Germany
\and
Institut f\"ur Astrophysik, Georg-August-Universit\"at G\"ottingen, Friedrich-Hund-Platz~1, 37077~G\"ottingen, Germany
\and
Center for Space Science, NYUAD Institute, New York University Abu Dhabi, Abu Dhabi, UAE
\and
Department of Physics, University of Warwick, Gibbet Hill Road, Coventry CV4 7AL, UK
\and
Centre for Exoplanets and Habitability, University of Warwick, Gibbet Hill Road, Coventry CV4 7AL, UK
\and
Zentrum f\"ur Astronomie und Astrophysik, TU Berlin, 
Hardenbergstra{\ss}e 36,
D-10623 Berlin, Germany
\and
Observatoire de Gen\`eve, Universit\'e de Gen\`eve, Chemin des maillettes 51, 1290 Sauverny, Switzerland
\and
Armagh Observatory \& Planetarium, College Hill, Armagh, BT61 9DG, UK
}

\date{Received; accepted}

 
  \abstract
   {
    The ESA PLAnetary Transits and Oscillations of stars (PLATO) mission will search for terrestrial planets in the habitable zone of solar-type stars.
    Because of telemetry limitations, PLATO targets need to be pre-selected.     
   }
   {
    In this paper, we present an all sky catalogue that will be fundamental to selecting
    the best PLATO fields and the most promising target stars, deriving their basic parameters, analysing the instrumental performances, and then planing
    and optimising follow-up observations. This catalogue also represents a valuable resource for the general definition of stellar samples optimised for
    the search of transiting planets.     
   }
   {
    We used \emph{Gaia} Data Release 2 (DR2) astrometry and photometry and 3D  maps of the local interstellar medium
    to isolate FGK (V$\leq$13) and M (V$\leq$16) dwarfs and subgiant stars.
   }
   {
    We present the first public release of the all-sky PLATO Input Catalogue
   (asPIC1.1) containing a total of 2~675~539 stars including 2~378~177 FGK dwarfs and subgiants and 297 362 M dwarfs. The median distance in our sample is 428~pc for FGK stars and 146 pc for M dwarfs, respectively. We derived the reddening of our targets and developed an algorithm to estimate stellar
    fundamental parameters (T$\rm_{eff}$, radius, mass) from astrometric and photometric measurements.
   }
   {
    We show that the overall (internal+external) uncertainties on the stellar
    parameter determined in the present study are $\sim$230 K (4$\%$) for the effective temperatures, $\sim$0.1 R$\rm_{\odot}$ (9$\%$) for the stellar radii, and
    $\sim$0.1 M$\rm_{\odot}$ (11$\%$) for the stellar mass. We release a special target list containing all known planet hosts cross-matched with our
    catalogue. 
   }

             
   \titlerunning{The all sky PLATO Input Catalog}
   \authorrunning{Montalto et al.} 
   \maketitle

\section{Introduction}

The PLATO mission \citep{rauer2014,rauer2016} is the third medium-class mission in the ESA Cosmic Vision programme. Its ambitious goals are the detection and characterisation of terrestrial planets around solar-type stars as well as the study of the properties of host stars. The focus of PLATO is on long-orbital-period terrestrial planets that are also in the habitable zone of solar-type stars. PLATO is designed for the discovery of Earth-analogue planets under potentially favourable conditions for the development of life.
   
PLATO will achieve its goals by detecting the low-amplitude dips in stellar brightness produced by planets transiting in front of the disk of their parent stars. 
The mission is designed and optimised to continuously monitor two Long-duration Observation Phase (LOP) fields, each one for up to three years, and has the capability to monitor some additional fields during shorter times, up to three months each (`Step and stare' Observation Phase, SOP). The precise observing strategy will be decided at a later stage (two years before launch at the latest), taking advantage of the advances in the field at that time.

PLATO will employ an array of 26 cameras
(hereafter `cameras' signifies the telescope optics and the focal plane assembly, including all the ancillary devices like baffling, electronics, etc.)
made up with refractive 20~cm-class optical elements with a very wide field (about 40 degrees in diameter) and a 12cm diameter equivalent aperture.

Each camera field of view is almost covered by an array of detectors totaling 1037 square degrees,  and 24 cameras are arranged in four groups of six cameras each, co-aligned to four slightly overlapping fields of view. The S/N of each star will therefore be dependent on its precise location within the overall PLATO field of view.
The combined field of view of the system of cameras amounts to a total of about 2100 square degrees \citep[][]{ragazzoni2015,magrin2018}.
The remaining two cameras are optimised for fast monitoring of very bright stars, their colour measurements \citep[e.g.][]{grenfell2020}, and
fine guidance and navigation. Their focal plane is equipped with frame transfer detectors that allow for a coverage somewhat larger than 600 degrees. The geometry of the covered patches in the sky is kept symmetric, under nominal conditions, for 90 degree rotations around the line of sight to keep the same coverage of the field of view throughout a pointing and to allow continuous in-flight calibration among different cameras.
Compliance with the performance specifications is computed using models for the degradation of the performances over the mission lifetime, mainly due to a moderate degradation of the throughput, due to the choice of rad-hardened glasses for the optical elements most exposed to cosmic radiation
\citep{magrin2016,corso2018}, and to the worst scenario in which  the detectors or electrical systems of two of the
cameras could fail by the End Of Life (EOF).
  
\noindent
Because of limits on the telemetry rate from the satellite to Earth, it is not possible to download full CCD images at a sufficient rate to detect planets; therefore, 
it is necessary to pre-select PLATO targets. For a subsample of targets, customised `postage stamps' (imagettes) are downloaded to ground; for the remaining selected stars, centring and photometry will be performed on board and then downloaded to ground.
Other, previous space missions followed the same approach. The \emph{CoRoT} Exo-Dat catalogue \citep{deleuil2009},
the \emph{Kepler} Input catalogue \citep[KIC;][]{brown2011},
the \emph{K2} Ecliptic Plane Input Catalogue
\citep[EPIC;][]{huber2016}, and
the \emph{TESS} Input Catalogue
\citep[TIC;][]{stassun2018,stassun2019}
were produced to select the best targets for exoplanet searches. 

In this paper, we present the all-sky PLATO Input Catalogue (asPIC), whose main purpose is to serve as the primary database for the selection of PLATO targets. Establishing the properties of PLATO host stars is crucial to determining the properties of transiting planets; for example, the uncertainty on their radii is directly related to the uncertainty on the host-star radius.  The classical properties of planet hosts are also essential to modelling the stars using asteroseismology.
Also, the habitability of a given planet depends on the properties of the host star, among other factors. Moreover, knowing the values of stellar parameters permits the statistical study of the frequency of planets as a function of stellar type, metallicity, environment, and so on \citep[e.g.][]{howard2012,bryson2020}.
   
\noindent
The construction of such a catalogue not only requires the identification, across the entire sky, of stars with the required spectral type and luminosity class, but also requires that noise constraints be accounted for to facilitate the detection of transiting planets around the selected sources. Hence limits on the magnitude range of the stellar samples are set according to mission requirements and the degree of contamination of each target from neighbouring sources must be evaluated. The PLATO pixel scale will be 15 arcsec pix$^{-1}$ on axis, which lies between that of \emph{Kepler} \citep[4 arcsec pix$^{-1}$; ][]{borucki2010} 
and that of \emph{TESS} \citep[21 arcsec pix$^{-1}$; ][]{ricker2014}. 

The optical quality of the PSF, although slightly variable over the field of view and depending on the thermal status of the cameras, is expected to encompass an area of a few pixels \citep[see also][]{gullieuszik2016,umbriaco2018}.
Consequently, it is expected that several sources, especially in crowded fields,  will be blended into a single brightness measurement area and they may become the cause of false positive signals 
\citep[e.g.][]{santerne2015,fressin2013}. 
More specifically, the asPIC will be used to: 
\begin{enumerate}
\item 
select the optimal PLATO observing fields; 
\item
select all FGKM dwarf and subgiant stars satisfying the magnitude and signal-to-noise constraints defined in the PLATO stellar sample requirements;
\item
estimate basic stellar parameters for all targets, such as temperatures, radii, and masses; 
\item identify known variable stars, binaries, members of multiple systems, active stars 
with bitmasks, and summarize the information available from existing catalogues, that is, mainly {\it Gaia} and TESS releases;
\item
supply a list of all target contaminants up to a specific angular distance from the target and up to a limiting magnitude, for the targets in the fields to be observed,
in order to understand their impact on photometric performances and because of the challenge that background contaminants pose to planetary candidate validation;
\item
guide the organisation and optimisation of ground-based follow-up strategies. 
\end{enumerate}
   
\noindent
Previous attempts to produce this catalogue showed that this is not a trivial task. In particular, it is hard to robustly discriminate between dwarfs and evolved stars without accurate knowledge of the  absolute luminosities of sources. Reduced proper motions were employed in the past to this purpose \citep{nascimbeni2016}, but this technique is now superseded by the availability of trigonometric parallaxes from \emph{Gaia} DR2 \citep{gaia2016,gaiaDR22018}\footnote{ 
Although \emph{Gaia} EDR3 has now been released \citep{gaia2021EDR3} this work is based on \emph{Gaia} DR2.
Future updates of the asPIC will be based on \emph{Gaia} DR3.
},  
essentially for all targets of PLATO interest, permitting accurate determination of distances and therefore of absolute luminosities. 
   
\noindent
After ESA adoption (in June 2017), PLATO recently entered the full industrial development phases C-D (March 2020)
to be ready for launch in 2026.  At this development stage, the selection of the PLATO
LOP fields is not yet finalised (it will be two years before launch). The present work is focused on the
asPIC. This catalogue is made publicly available to the community and it will be updated whenever relevant changes occur, including new {\it Gaia} releases.
 
\noindent
After briefly describing the mission requirements defining PLATO stellar samples in Sect.~\ref{sec:stellar_samples}, we describe the target selection criteria in Sect.~\ref{sec:selection_criteria}.
Then, in Sect~\ref{sec:reddening}, we describe how we account for reddening and absorption in our analysis. Our stellar parameter pipeline is presented in Sect.~\ref{sec:stellar_parameters}.
In Sect.\ref{sec:comparisons}, we compare our results with those published in various literature sources. In Sect.~\ref{sec:special_list}, we
describe a special list of known planet host stars included in the catalogue release.
Section~\ref{sec:conclusions} provides a brief summary of the present work.

\begin{table*}
\caption{Summary of science requirements for the PLATO stellar samples.
\label{tab:samplesrequirements}}
\centering                          
\begin{tabular}{lccccc}        
\hline                 
  & & & & & \\
  & P1 & P2 & P4 & P5 & Colour sample \\
  & & & & & \\
\hline                        
 & & & & & \\
Stars & $\ge$15 000 (goal 20 000) & $\ge$1000 & $\ge$5000 & $\ge$245 000 & 300 \\
 & & & & & \\
Spectral Type & Dwarf and  &  Dwarf and &  M Dwarfs &  Dwarf and  & Anywhere in \\
 & subgiants F5-K7 & subgiants F5-K7 &  & subgiants F5-late K & the HR diagram \\
  & & & & & \\
Limit {\it V} & 11 & 8.5 & 16 & 13 & - \\
 & & & & & \\
Random noise (ppm in 1 hr) & $<$50 & $<$50 & - & - & - \\
 & & & & & \\
Wavelength (nm)& 500--1000 & 500--1000 & 500--1000 & 500--1000 & Blue (500--675)\\
 & & & & & and\\
 & & & & & red (675--1000)\\
 & & & & & spectral bands\\
 & & & & & \\ 
\hline                                   
\end{tabular}
\end{table*}

\section{PLATO stellar sample requirements}
\label{sec:stellar_samples}

The  four main PLATO stellar samples (named P1, P2, P4, and P5)\footnote{For historical reasons, the sample P3 has been eliminated, but the numbering of the PLATO samples was left unchanged.} will be observed in the wavelength range between 500 nm and 1000 nm, and an additional colour sample will be observed in two broad blue and red spectral bands spanning the 500--675 and 675--1000 nm wavelength ranges, respectively \citep{corso2018}.
The science requirements for the PLATO stellar samples are listed below and summarised in Table~\ref{tab:samplesrequirements} according to the ESA Science Requirements Document (SciRD, PTO-EST-SCI-RS-0150\_SciRD\_7\_0).
The definition of the PLATO samples requires knowledge of the visual apparent magnitude $V$
because the SciRD adopts this magnitude as reference to identify sufficiently
bright targets for ground-based spectroscopic follow-up. 
The \textit{V} magnitude we used to select the targets comes from our own calibrated transformation of \textit{Gaia} DR2 magnitudes and colours to the Johnson \textit{V} band (Eq.~\ref{eq:color_transformation}). This choice ensures the maximum homogeneity 
of the \textit{V} band photometry.
The calibration procedure is described in Appendix~\ref{sec:gaiaV}.
The colour transformation relation is 

\begin{align}\label{eq:color_transformation}
(G-V)_0 & =c_0+c_1 (G_{\textrm{BP}}-G_{\textrm{RP}})_\textrm{0} + 
           c_2(G_{\textrm{BP}}-G_{\textrm{RP}})^2_0 + \nonumber \\
        & + c_3(G_{\textrm{BP}}-G_{\textrm{RP}})^3_0 + c_4(G_{\textrm{BP}}-G_{\textrm{RP}})^4_0 + \nonumber \\
        & + c_5(G_{\textrm{BP}}-G_{\textrm{RP}})^5_0,
\end{align}

\noindent
where {\it G}, $G_{\textrm{BP}}$, $G_{\textrm{RP}}$ are the {\it Gaia} DR2 magnitudes
and the null subscript indicates
colours dereddenend with the procedure described in Sect.~\ref{sec:reddening}.
The best-fit coefficients of the transformation relation are reported in Table~\ref{tab:coeffpolgvbp}.

\subsection{Stellar sample 1 (P1)}
\begin{itemize}
    \item The total number of targets in stellar sample 1 (cumulative over all sky fields) shall be at least 15 000 dwarf and subgiant stars of spectral types from F5 to K7, with a goal of 20 000.
    \item The dynamic range of stellar sample 1 shall be {\it V} $\le$ 11.
     \item The random noise level for stellar sample 1 shall be below 50 ppm in 1 h.
    \item In stellar sample 1, the proportion of brighter targets ({\it V} $\le$ 10.5) shall be maximised.
    \item Stellar sample 1 shall be observed during a LOP.
\end{itemize}

\subsection{Stellar sample 2 (P2)}
\begin{itemize}
\item The total number of targets in stellar sample 2 (cumulative over all sky fields) shall be at least 1000 dwarf and subgiant stars of spectral types from F5 to K7.
\item The dynamic range of stellar sample 2 shall be {\it V} $\le$ 8.5.
\item The random noise level for stellar sample 2 shall be below 50 ppm in 1 h.
\item Stellar sample 2 shall be observed during a LOP.
\end{itemize}

\subsection{Stellar sample 4 (P4)}
\begin{itemize}
\item The total number of targets in stellar sample 4 (cumulative over all sky fields) shall be at least 5000 cool late-type dwarfs (M dwarfs) monitored during a LOP.
\item The dynamic range of stellar sample 4 shall be {\it V} $\le$ 16.
\end{itemize}

\subsection{Stellar sample 5 (P5)}
\begin{itemize}
\item The total number of targets in stellar sample 5 (cumulative over all sky
fields) shall be at least 245 000 dwarf and subgiant stars of spectral types
from F5 to late K.
\item The dynamic range of stellar sample 5 shall be {\it V} $\le$ 13.
\item Stellar sample 5 shall be observed during a LOP.
\end{itemize}

\subsection{Colour sample}
\begin{itemize}
\item Part of the payload must provide photometry in at least two separate
colour broad-bands.
\item PLATO shall have the capability to observe 300 stars (in at least two
pointings) located anywhere on the HR diagram, with colour information,
imagettes, and a sampling time equal to 2.5 s.
\end{itemize}

The above stellar samples  are defined only
across the two selected PLATO LOP fields. The final decision on these fields will be taken two years before launch. So far, two preliminary LOP fields have been identified.
The P1 sample is a subsample of P5. However, the  P1 sample is also S/N-limited and, because of the particular configuration of PLATO pointing, it can be defined only once the LOP fields are selected. 

In this work, we construct an all-sky catalogue, focusing on  only two main PLATO stellar samples. One sample is related to FGK (limited to F5) dwarf and subgiant stars with apparent visual magnitude $V\leq13$ (and formally corresponds to the P5 sample as defined above), while the other sample is related to M-dwarfs with $V\leq16$ (which corresponds to P4). For the sake of clarity, hereafter we refer to these two samples as the FGK sample and the M sample, respectively. 
We note that although
the asPIC is an all-sky catalogue,
PLATO will not point towards the Galactic plane.

\begin{table*}
\caption{Coefficients of the polynomial best-fit model relating the
$(G-V)_0$ and ({\it G$\rm_{BP}$-G$\rm_{RP}$})$_0$ colours and
residuals of the fit (see also Appendix~\ref{sec:gaiaV}). The relation is valid within 
0.5$<$({\it G$\rm_{BP}$-G$\rm_{RP}$})$\rm_0<$5.
\label{tab:coeffpolgvbp}}
\centering                          
\begin{tabular}{cccccccc}        
\hline                 
  &  & ($G_{\textrm{BP}}$-$G_{\textrm{RP}}$)$_\textrm{0}$ & ($G_{\textrm{BP}}$-$G_{\textrm{RP}}$)$_\textrm{0}^2$ & ($G_{\textrm{BP}}$-$G_{\textrm{RP}}$)$_\textrm{0}^3$ & ($G_{\textrm{BP}}$-$G_{\textrm{RP}}$)$_\textrm{0}^4$ & ($G_{\textrm{BP}}$-$G_{\textrm{RP}}$)$_\textrm{0}^5$ & $\sigma\rm_{fit}$\\    
   & $c_0$ & $c_1$ & $c_2$ & $c_3$ & $c_4$ & $c_5$  \\    
\hline                        
(\textit{G}-\textit{V})$_\textrm{0}$ &    -0.17276 &      0.47885 &     -0.71953 &      0.24374 &     -0.04458 &  0.00317 & 0.02745\\
\hline                                   
\end{tabular}
\end{table*}

\section{Selection criteria}
\label{sec:selection_criteria}

The selection of the PLATO targets was performed using the
absolute colour-magnitude diagram (CMD) corrected for reddening.
First, we employed theoretical models based on Galactic simulations to identify the region of the diagram where our targets are located (Sect.~\ref{sec:theoretical_models}). On the same diagram, we then represented  stars 
that, based 
on available spectroscopic and/or astrometric observations, have the properties required by the SciRD 
(Sect.~\ref{sec:control_sample}). Finally, by taking into account both the theoretical and empirical distributions of stars in the
CMD, we proceeded by defining an analytical selection
region,
and estimated its degree of completeness and contamination (Sect.~\ref{sec:analytic_approximation}). 
Finally, in Sect.~\ref{sec:asPIC1.1} we show
that our selection does not imply any bias in the metallicity distribution of the selected stars with respect to a sample of local stars (Sect.~\ref{sec:selection_metal}) and 
present
the  asPIC1.1  catalogue resulting from our selection. 

As summarised in Sect.~\ref{sec:stellar_samples}, the
PLATO samples considered here include an FGK dwarf and subgiant sample (limited to F5), and an M dwarf
sample. Considering the definitions reported in Table~5 of \cite{pecaut2013}, spectral type F5 corresponds to T$_\textrm{eff} = 6510\,$ K, or to an unreddened {\it Gaia} colour 
({\it G$\rm_{BP}$-G$\rm_{RP}$})$_0$ = 0.587, and spectral type M0 corresponds to T$_\textrm{eff} = 3870\,$K or unreddened ({\it G$\rm_{BP}$-G$\rm_{RP}$})$_0$ = 1.84, where {\it G$\rm_{BP}$} and {\it G$\rm_{RP}$} are the blue and red {\it Gaia} magnitudes, respectively. 
In addition, dwarfs and subgiants have $\log \textrm{g} > 3.5$.

\subsection{Theoretical models}
\label{sec:theoretical_models}

\begin{figure*}
        \centering
        \includegraphics[width=0.48\textwidth]{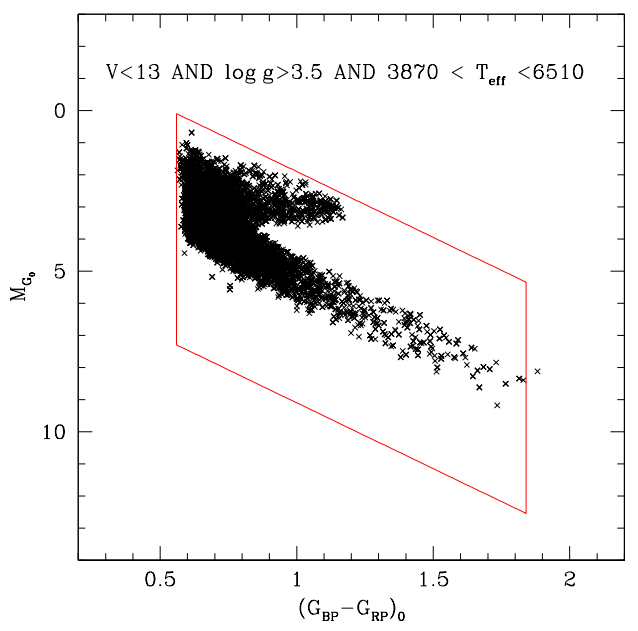}
        \includegraphics[width=0.48\textwidth]{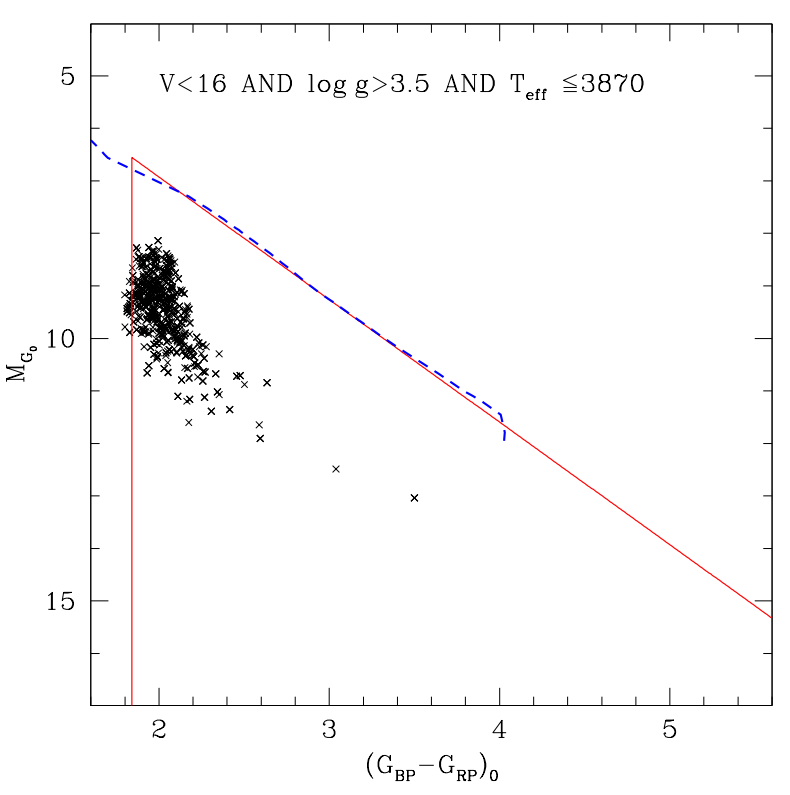}
        \caption{Absolute magnitude vs. reddening-free colour obtained from TRILEGALv1.6 simulations.
        Crosses represent simulated stars from a total area of 100 square degrees in the NPF and the SPF, satisfying the conditions described by the labels at the top of each plot. The left and right panels also show the analytical selection (red lines) described in Sect.~\ref{sec:analytic_approximation} for FGK
        dwarfs and subgiants (FGK sample) and for M dwarfs (M sample), respectively. In the diagram on the right, the blue dashed line represents a 10 Myr solar metallicity isochrone from the Padova database \citep{bressan2012}.
        }
        \label{fig:trilegal}
\end{figure*}

\begin{figure*}
        \centering
        \includegraphics[width=0.3\textwidth]{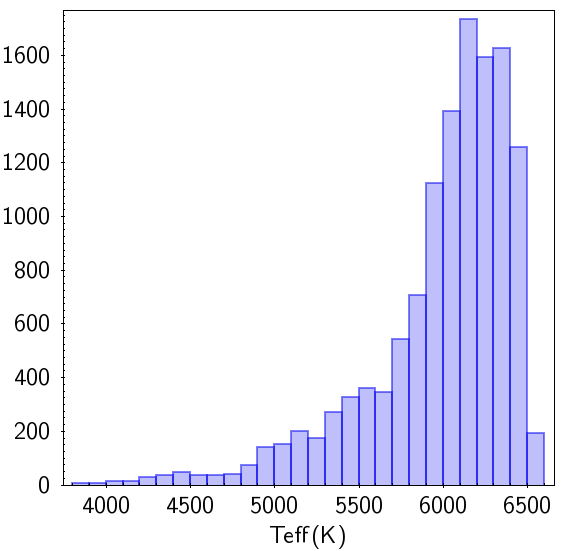}
        \includegraphics[width=0.3\textwidth]{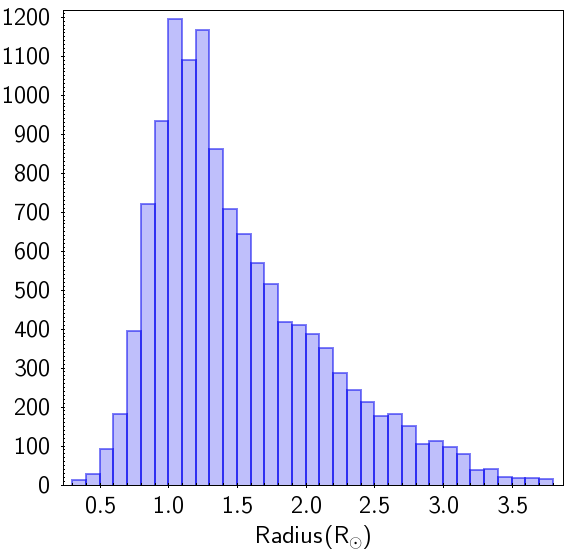}
        \includegraphics[width=0.3\textwidth]{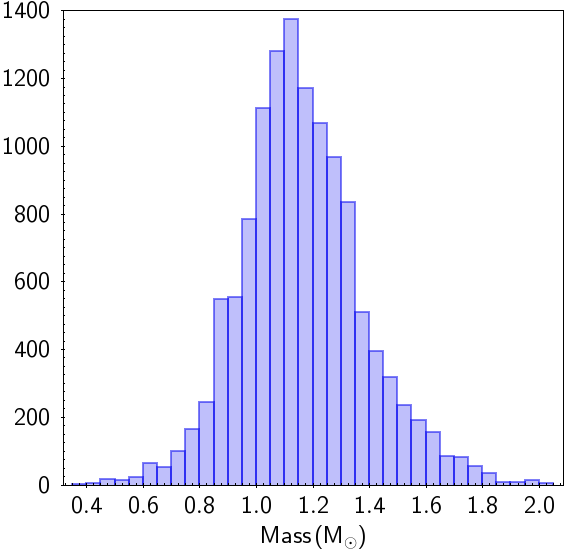}
        \caption{Distribution of effective temperatures (left panel), stellar radii (middle), and masses (right)
                 for FGK sample stars from TRILEGAL simulations.}
        \label{fig:teffdist_trilegal}
\end{figure*}

\begin{figure*}
        \centering
        \includegraphics[width=0.3\textwidth]{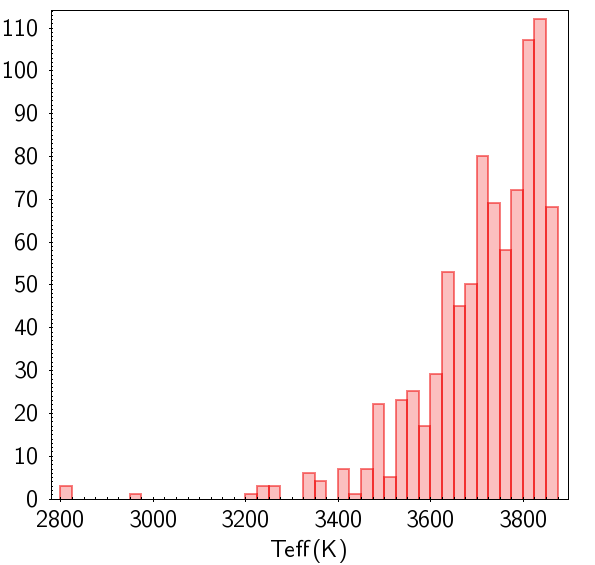}
        \includegraphics[width=0.3\textwidth]{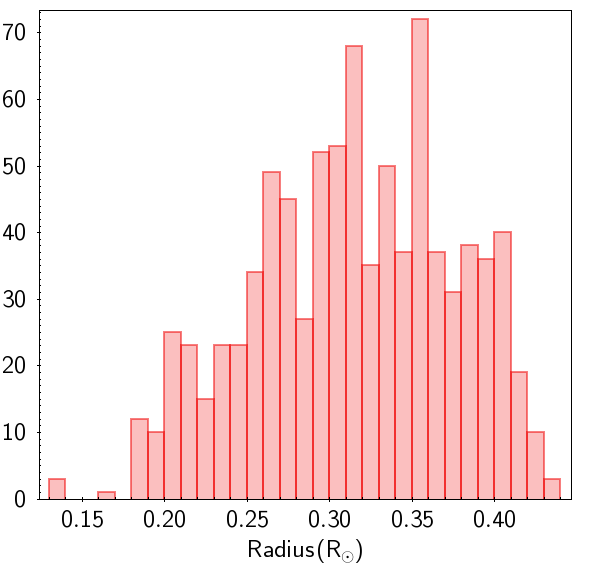}
        \includegraphics[width=0.3\textwidth]{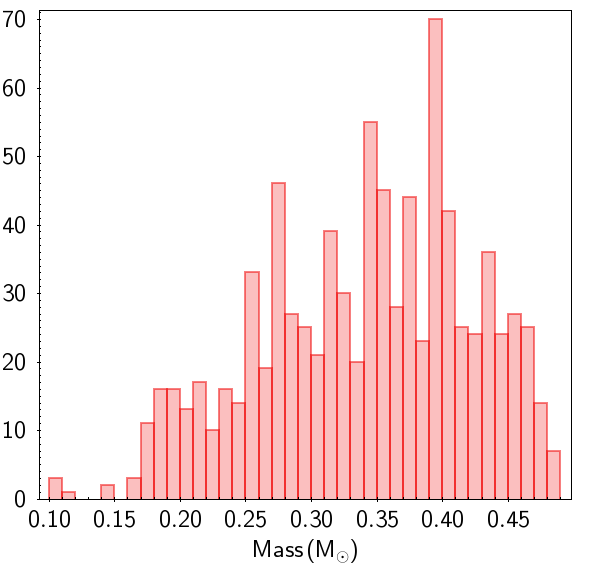}
        \caption{Distribution of effective temperatures (left panel), stellar radii (middle), and masses (right) for M sample stars from TRILEGAL simulations.}
        \label{fig:teffdist_trilegal_P4}
\end{figure*}

To determine and calibrate our selection criteria in an observational CMD, we used Galactic simulations
from TRILEGALv1.6,  \citep{girardi2005}\footnote{http://stev.oapd.inaf.it/cgi-bin/trilegal}. 
TRILEGAL is a population synthesis code for simulating the stellar photometry of any Galaxy field. It allows the user to simulate the photometry in several different broad-band photometric systems.
Each simulation we performed corresponds to an area of ten square degrees. The simulations were centred at ({\it l,b})=(65$^{\circ}$, 30$^{\circ}$) and ({\it l,b})=(253$^{\circ}$, -30$^{\circ}$), and on four surrounding locations ten degrees apart from the central fields in Galactic longitude or latitude. The two central fields correspond to the two provisional PLATO long-duration fields in the Northern and Southern Galactic hemispheres, that is the North PLATO field and the South PLATO field
(NPF and SPF, respectively). We considered the \textit{Gaia} DR2 passbands of \citet{appelaniz2018} and determined the apparent visual magnitude \textit{V} using the relation reported in Appendix~\ref{sec:gaiaV}.
We corrected the apparent magnitudes of each simulated star for extinction using the {\it V} band extinction tabulated in the simulations and the relations given in Sect.~\ref{sec:conversion} in order
to convert the extinction in the {\it V} band to the extinction in the {\it Gaia} bands.
Figure~\ref{fig:trilegal} shows the intrinsic colour (reddening free) and absolute magnitude obtained from TRILEGALv1.6 simulations. 
The left panel shows stars
with \textit{V}$<$13, log$\,$g$>$3.5 and 3870~K~$<\textrm{T}_{\textrm{eff}}<6510$~K corresponding to
our adopted limits for the FGK sample, while the right panel shows stars with \textit{V}$<$16, log$\,$g$>$3.5 and $\textrm{T}_{\textrm{eff}}\leq 3870$~K (M sample).
Figures~\ref{fig:teffdist_trilegal} and~\ref{fig:teffdist_trilegal_P4} show the distributions of effective temperatures (left panel) radii (middle) and masses (right) of the simulated stars. Magnitude limited samples, like the ones we are considering, are dominated by hot-dwarfs and subgiants, because these stars are intrinsically more luminous and therefore observable in a large volume.

We also repeated the simulations for fields located in the Galactic plane or at the Galactic poles. The intrinsic colours and magnitudes of the selected stars are perfectly compatible with the ones shown in Figure~\ref{fig:trilegal}. 
Larger uncertainties in the reddening estimates occur
for the Galactic plane and would lead to higher levels
of contamination, particularly for hot dwarfs and subgiants
(see also Sect.~\ref{sec:analytic_approximation}).

\subsection{Empirical control samples}
\label{sec:control_sample}

\begin{figure*}
        \centering
        \includegraphics[width=0.49\textwidth]{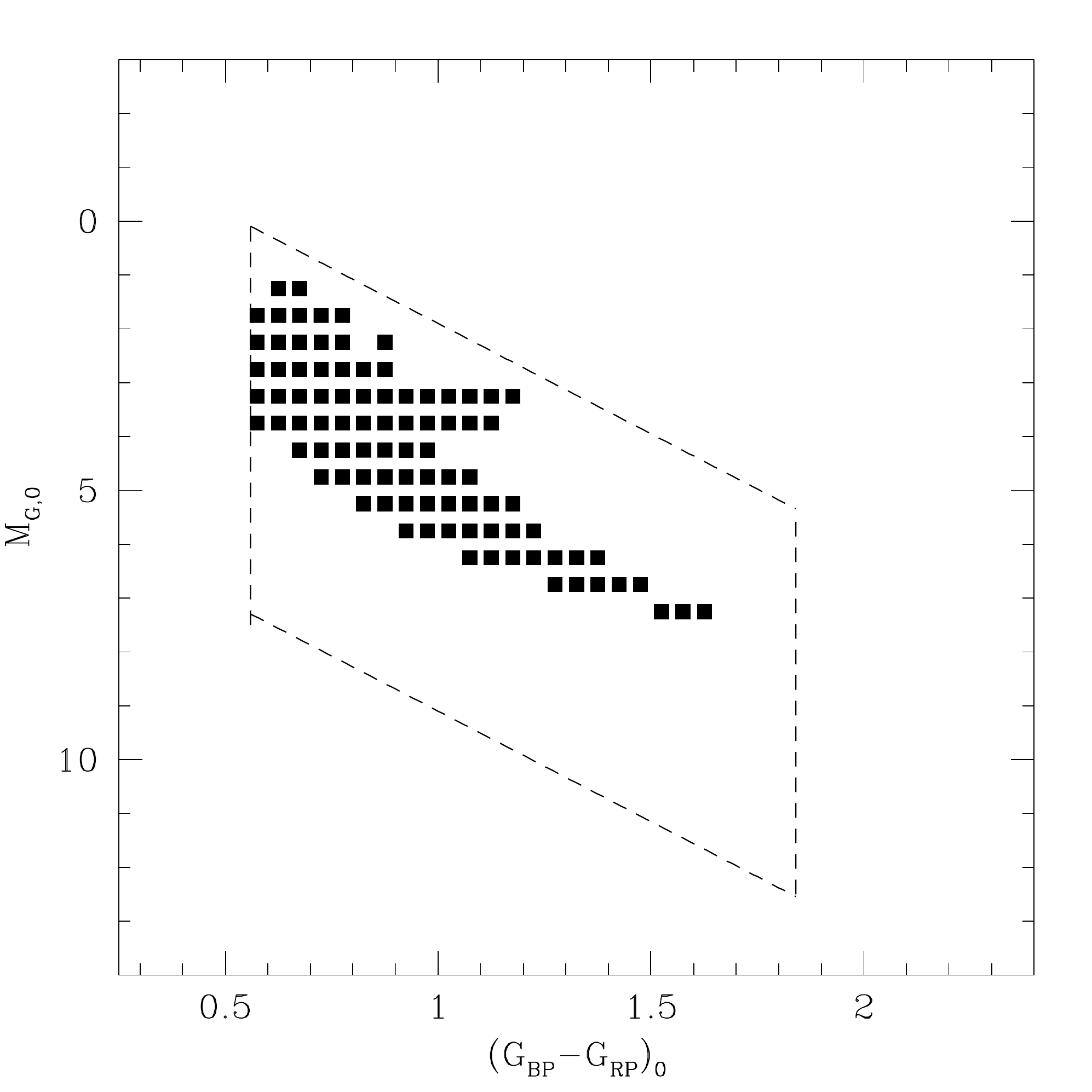}
    \includegraphics[width=0.47\textwidth]{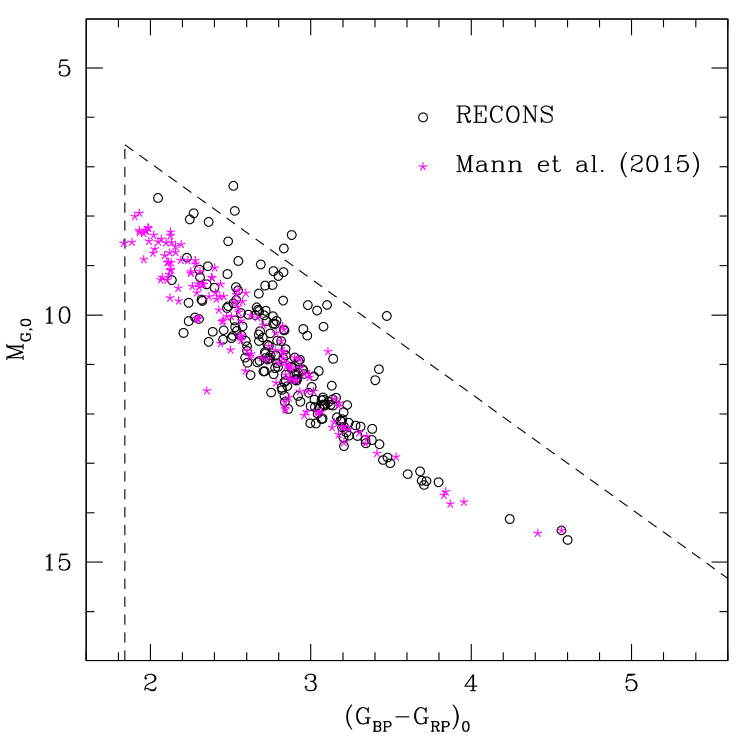}
        \caption{
        In the left panel, black rectangles represent FGK dwarfs and subgiants from RAVE \citep[see Sect.~\ref{sec:control_sample},][]{kunder2017}.
        The right panel shows M-dwarfs \citep[open circles]{henry2018} and \citet[magenta asterisks]{mann2015} catalogues.
        Dashed lines in both panels show the analytical selection
        defined in Sect.~\ref{sec:analytic_approximation}
        for the FGK (left) and M stellar samples (right).}
        \label{fig:control_sample}
\end{figure*}

For FGK dwarfs and subgiants, as an empirical control sample, we used stars from the RAVE DR5 survey \citep{kunder2017}. RAVE provides two sets of stellar parameters: those that are direct outputs of the stellar parameter pipeline and the calibrated parameters. Because the direct outputs are based on a grid of synthetic spectra which may not match real spectra over the entire domain of the parameter space, a calibration based on reference stars is performed in order to minimise systematic offsets. For example, surface gravities
are calibrated using both the asteroseismic $\log g$ values of 72 giants from \cite{valentini2016} 
and the \textit{Gaia} benchmark dwarfs and giants \citep{heiter2015}. We then adopted the calibrated parameters.
We considered stars  with RAVE parameter ALGOCONV=0 and $S/N>50$ to ensure the good quality of extracted parameters.
A total of 122 167 stars were used. We analysed the distribution of RAVE stars in the absolute colour magnitude diagram constructed using \cite{bailer2018} distances. We corrected the apparent magnitudes for extinction using the procedure presented in Sect.~\ref{sec:reddening}
\footnote{We also compared our reddening {\it E(B-V)} with the one calculated in RAVE and found perfectly consistent results: 
$\Delta${\it E(B-V)}=(-0.01$\pm$0.02).}. To identify the selection regions of dwarfs and subgiants in the ($G_{\textrm{BP}}-G_{\textrm{BP}}$)$_0$, $M_{G,0}$  plane, we considered a grid of $0.05\, \textrm{mag} \times 0.5\, \textrm{mag}$ bins in colour and magnitude across the plane. We considered stars with \textit{V}$<$13, where the \textit{V} band magnitude was calculated with the calibration relationship reported in Appendix~\ref{sec:gaiaV}. Stars for which RAVE parameters did not satisfy the conditions log g$>3.5$ and 3870 K$<T_{\textrm{eff}}<$6510 K were considered contaminants. We then selected those grid rectangles with at least 50 stars and for which the number of dwarfs and subgiants was larger than $75\%$ of the total number of stars. Figure~\ref{fig:control_sample} (left panel) shows the result of this procedure. The black rectangles are the regions belonging to the dwarf+subgiant class.

For M-dwarfs, we used 
The {\it REsearch Consortium On Nearby Stars}  (RECONS\footnote{www.recons.org}) catalogue
 \citep{henry2018}, and the \citet{mann2015} catalogue.
Among other information, the RECONS catalogue reports the apparent visual Johnson magnitude of the targeted stars. Whenever multiple measurements for the same object were present in the catalogue, we averaged their visual magnitudes.
From the initial list of 348 entries we obtained a list of 289 measurements for the individual objects.
The catalogue was cross-matched with \textit{Gaia} DR2 and
we then selected stars with $V<16$ that  satisfy Eq.~(1) and Eq.~(2) of \cite{arenou2018} \textit{Gaia} astrometric and photometric quality criteria and are classified as M-dwarfs\footnote{RECONS does not report the effective temperature, but it reports the spectral type of the stars.}, restricting the list to 195 stars. We then calculated the reddening as described in Sect.~\ref{sec:reddening}. The right panel of Figure~\ref{fig:control_sample} shows the absolute colour--magnitude diagram corrected for reddening of the selected stars (open black circles) of this sample. \\
\noindent
\cite{mann2015} constructed a sample of nearby late K and M dwarfs
(all within 40 pc from the Sun) for which optical and near-infrared spectra were obtained. We used the effective temperatures derived by these latter authors to isolate only M dwarfs
by
imposing T$_{\textrm{eff}}\leq3870$ K. \textit{Gaia} DR2 was cross-matched with this catalogue and we restricted the sample to those stars satisfying Eq.~(1) and Eq.~(2) of \citet{arenou2018}. The apparent visual magnitude in the Johnson system (necessary for our selection, and not reported in the catalogue) was calculated as described in Appendix~\ref{sec:gaiaV}. By means of these selection criteria the initial sample of 179
stars was reduced to 153 stars represented by the magenta asterisks in Figure~\ref{fig:control_sample}.

\subsection{Analytical selection}
\label{sec:analytic_approximation}

\begin{figure}
        \centering
        \includegraphics[width=9cm]{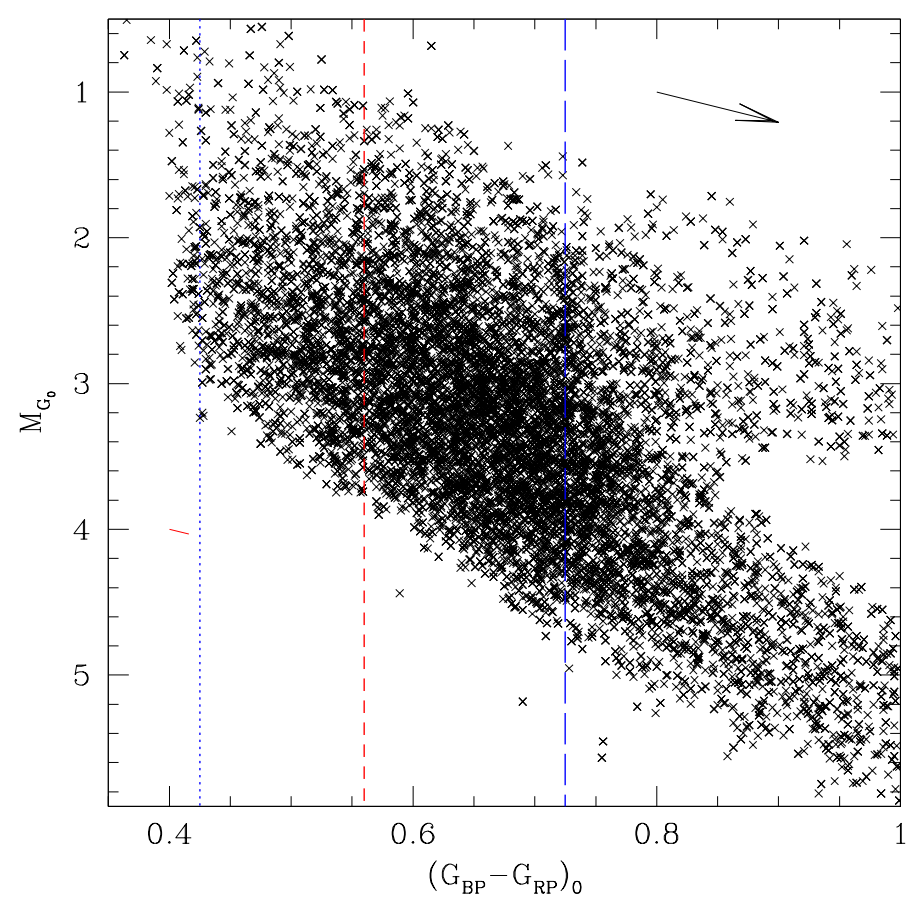}
        \includegraphics[width=9cm]{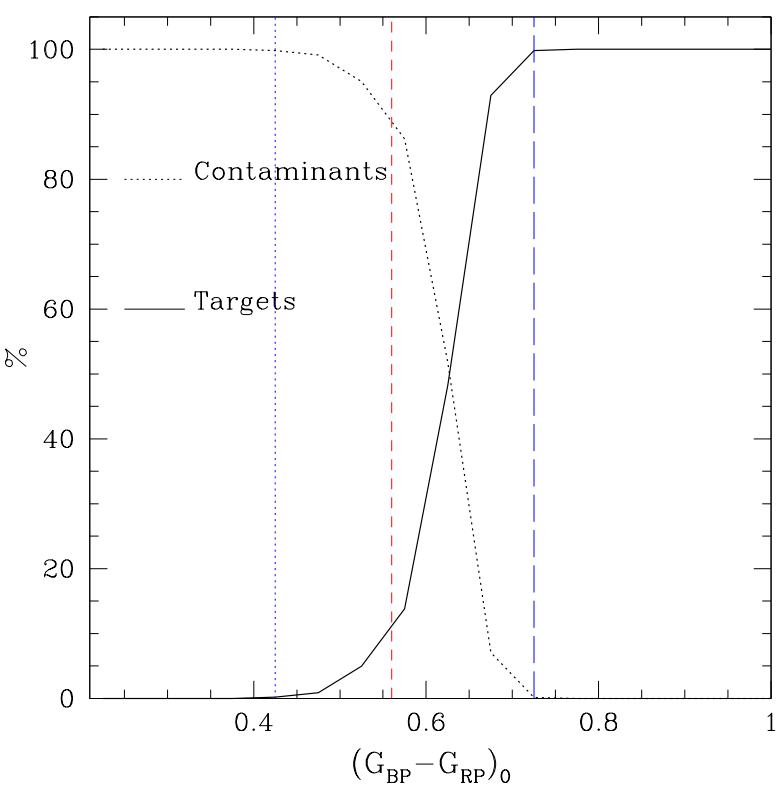}
        \caption{
                 Top: Colour--magnitude diagram from Galactic simulations presenting
                 dwarfs and subgiant
                 stars with {\it V}$<$13
                 and T$\rm_{eff}<$7220 K (i.e. later than F0).
                 The arrow indicates the average direction of the reddening vector.  
                 Bottom: Percentage of targets (continuous line) and contaminants (i.e. stars with spectral type earlier than F5, dotted line) within 0.05 mag wide colour intervals. The red dashed vertical lines denote our adopted blue colour limit of the PLATO FGK sample,
                 the blue dotted vertical lines indicate the colour where the selection is 100\% dominated by contaminants, while the blue long-dashed vertical lines show the colour where the selection is 100\% dominated by the targets.
        }
        \label{fig:colour_extension}
\end{figure}

For the selection of M and FGK targets from the 
{\it Gaia} CMD, we then proceeded by defining a
set of inequations.
The red colour limit corresponding to ({\it G$\rm_{BP}$-G$\rm_{RP}$})$_\textrm{0}$=1.84 was used to separate M-dwarfs from FGK dwarfs and subgiants following \citet{pecaut2013}, while the blue colour limit
of the FGK samples was set at ({\it G$\rm_{BP}$-G$\rm_{RP}$})$_\textrm{0}$=0.56 corresponding to a spectral type between F4V-F5V according to \citet{pecaut2013}.
This limit is bluer than the ({\it G$\rm_{BP}$-G$\rm_{RP}$})$_\textrm{0}$=0.587 proposed by \citet{pecaut2013} for F5 stars in order
to (partially) account for reddening errors.
By using TRILEGAL
simulations, we perturbed the reddening correction
according
to the expected reddening uncertainties derived from our map (see Sect.~\ref{sec:reddening}),
and estimated the number of contaminants (stars with spectral type earlier than F5) and targets
in narrow colour intervals of 0.05 mag. 
Because our samples are magnitude limited, hotter stars
are more numerous than cool 
ones,
because they are intrinsically more luminous (see e.g. Fig.~\ref{fig:teffdist_trilegal}, left panel). 
The major source of contamination comes from massive
dwarfs and subgiants of spectral type earlier than F5. 
Figure~\ref{fig:colour_extension} (top panel) shows a
colour--magnitude diagram from Galactic simulations presenting dwarfs and subgiant stars with {\it V}$<$13 and T$\rm_{eff}<$7220 K (i.e. later than F0).
At the limit of
$(G_{\textrm{BP}}-G_{\textrm{RP}})_0$=0.56 (dashed red line)
simulations show that we expect a loss of $\sim$10 \%\ of good targets,
but an inclusion of $\sim$90 \%\ of contaminating stars.
Extending the selection criteria to bluer colours would increase
the number of contaminating stars (i.e. stars with spectral type earlier than F5),
with negligible recovery of good targets
(bottom panel of Fig.~\ref{fig:colour_extension}).
The upper luminosity threshold of FGK subgiants was set to include
stars down to log g=$3.5$.
Such a definition
also
helps to reduce the contamination from evolved, reddened sources. The lower luminosity threshold of FGK dwarfs is set at about three magnitudes below the main sequence to avoid contamination from potentially spurious sources populating regions of the CMD where no plausible main sequence dwarf is expected to be found. 
With our selections, distant red giants are excluded by construction independently of their reddening and the same holds   for white dwarfs.

In general, theoretical
predictions
and  observational
data
agree reasonably well. For M-dwarfs, such as those
visible
in Figs.~\ref{fig:trilegal} and~\ref{fig:control_sample}, simulated stars appear systematically offset toward larger absolute magnitudes than observations, and
appear to
span a smaller colour range.  Some differences are also evident among the observational samples. Several RECONS stars are intrinsically brighter than the \cite{mann2015} sample stars. The upper absolute luminosity limit of the M sample was set by performing a linear regression of a 10 Myr solar metallicity isochrone from the Padova database \citep{bressan2012}, represented by the blue dashed line in Fig.~\ref{fig:trilegal}. Considering the theoretical and observational uncertainties, this 
threshold is a good compromise to
include all cool late-type dwarfs at any metallicity,
including binaries and
active stars. At the same
time, by adopting this limit we are more likely
to discard giant stars, which are expected to be brigher than this limit. 
To reduce the contamination from reddened stars lurking inside our selection region we also limited the distance of the stars in the M sample to 600 pc. 
The rationale behind this
choice is related to our limiting magnitude \textit{V}=16. Assuming the conversion between Johnson \textit{V}  and Gaia magnitudes in Appendix~\ref{sec:gaiaV}, 
the maximum distance at which we can detect an unreddened M0 type star with an apparent visual magnitude \textit{V}=16 and 
at the bright limit of our selection ($M_\textrm{G,0}$=6.55)  is 573 pc.
Therefore, any other M-type star {within} our selection region cannot be found at a distance of greater than 600 pc.  
In summary, our adopted selection criteria are defined as follows:

\begin{equation} 
    \textrm{M sample}=\begin{cases}
    (G_{BP}-G_{RP})_0 \geq 1.84 \\
    M_{G,0} > 2.334\,(G_{BP}-G_{RP})_0+2.259 \\
    \textrm{Distance}<600\quad\textrm{pc}\\
    V\leq16\label{eq:P4}
    \end{cases}
,\end{equation}

\begin{equation} 
    \textrm{FGK sample}=\begin{cases}
    0.56 \leq (G_{BP}-G_{RP})_0 < 1.84 \\
    M_{G,0} \leq 4.1\,(G_{BP}-G_{RP})_0+5.0 \\
    M_{G,0} \geq 4.1\,(G_{BP}-G_{RP})_0-2.2\\
    V\leq13\label{eq:P5}
    \end{cases}
.\end{equation}

\noindent
The quantities that appear in the relations above are the intrinsic colour ({\it G$\rm_{RP}$-G$\rm_{BP}$})$_0$ in the Gaia bands, the Gaia intrinsic absolute magnitude ($M_{G,0}$), the apparent visual magnitude ($V$), and the distance from \cite{bailer2018}. 
The intrinsic colour is determined by deredenning as discussed in Sect. \ref{sec:reddening}. The selection regions are illustrated
in Fig.~\ref{fig:trilegal} (red lines) and in Fig.~\ref{fig:control_sample} (dashed lines). \\
By adopting this analytic selection we find that $92\%$ of the stars whose RAVE parameters are compatible with the selection criteria are included. 
For the M-dwarf samples we find that 97$\%$ (190 out of 195 stars) of the \citet{henry2018} stars, and 99$\%$ (152 out of 153 stars) of the \citet{mann2015} stars are included in the selection.

An alternative estimate of the completeness and contamination can be obtained using theoretical models. We used the same simulations described above in this section to determine the true positive rate (TPR) and the false positive rate (FPR) implied by our selection. We distinguished the cases of M and FGK selected stars. A star is considered a true positive (TP) if it satisfies  
inequation~\ref{eq:P4} for the M sample or inequation~\ref{eq:P5} for the FGK sample and the corresponding theoretical constraints (
\textit{V}$<$16, log$\,$g$>$3.5, and $\textrm{T}_{\textrm{eff}}\leq 3870$ K for M sample and
\textit{V}$<$13, log$\,$g$>$3.5, and $3870$ K $<\textrm{T}_{\textrm{eff}}<6510$ K for FGK sample); it is a false negative (FN) if it does not satisfy inequation~\ref{eq:P4} (or inequation~\ref{eq:P5}) but does satisfy the theoretical constraints; it is a false positive (FP) if inequation~\ref{eq:P4} (or inequation~\ref{eq:P5}) is satisfied but the theoretical constraints are not; and finally, it is a true negative (TN) if both inequation~\ref{eq:P4} (or inequation~\ref{eq:P5}) and the theoretical constraints are not satisfied. The TPR and FPR are therefore defined as TPR=$\rm\frac{TP}{TP+FN}$ and FPR=$\rm\frac{FP}{FP+TN}$. 
The TPR and the FPR are measures of the performances of the selection algorithm. 
The TPR of the FGK selection is 100\% with a FPR of 12\%. The TPR of the M selection is 88\% with a FPR of 0\%. The FGK selection appears more `permissive' than the M
selection, implying higher completeness but also higher contamination. Most of the decrement in the TPR of the M sample appears to be related to the sharp dependence on metallicity of stellar models of late-type stars.

\begin{figure}
        \centering
        \includegraphics[width=9cm]{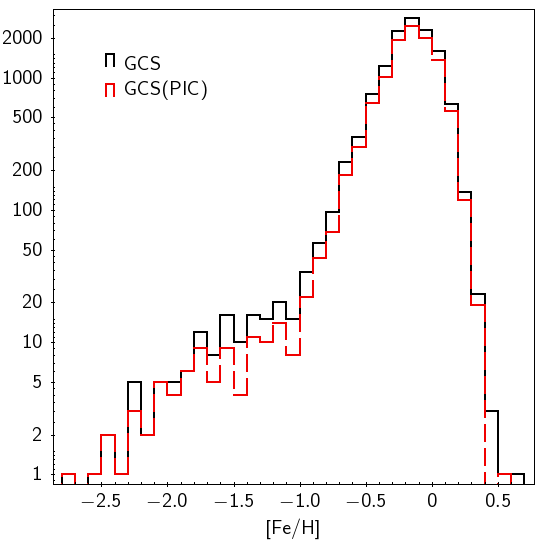}
        \caption{
    Metallicity distribution of stars selected from the GCS catalogue \citep{holmberg2009}. The black histogram shows all stars from the catalogue (with T$\rm_{eff}<$6500 K) while the red-dashed histogram shows all stars selected according to asPIC criteria (inequation~\ref{eq:P4} and inequation~\ref{eq:P5}).
        }
        \label{fig:GCS_met}
\end{figure}

\subsection{Impact of selection on the metallicity distribution}
\label{sec:selection_metal}

It is important to understand if the selection criteria we adopted introduced any appreciable bias on the metallicity distribution of the target stars. To this purpose, we retrieved the Geneva-Copenhagen Survey (GCS) catalogue \citep{holmberg2009} and cross-matched it with asPIC
by angular separation (accounting for proper motions) and brightness (by comparing the \textit{V} magnitude of GCS with the \textit{V} magnitude derived from \textit{Gaia}; see Appendix~\ref{sec:gaiaV}). In particular, we considered two samples, one with all stars in the GCS having T$\rm_{eff}<6500$ K and for which the value of the metallicity [Fe/H] was reported (in total 12 652 stars), and another one with all stars in the GCS for which the asPIC selection criteria (Eq.~\ref{eq:P4} and Eq.~\ref{eq:P5}) were satisfied (and had a metallicity estimate; in total 10 798 stars). We then compared the corresponding distributions of metallicities as shown in Fig.~\ref{fig:GCS_met}. We performed a Kolmogorov-Smirnov two-sided test to evaluate whether or not the two distributions were drawn from the same population. We consider a significance level equal to $\alpha=$0.01 and obtain a p-value equal to 0.6625.
The null hypothesis is therefore not rejected and the two distributions are statistically equivalent. This allows us to conclude that our  selection criteria are not biasing the distribution of metallicities of the target stars (with respect to the distribution of metallicities of the GCS catalogue). 

\subsection{asPIC1.1}
\label{sec:asPIC1.1}

\begin{figure}
        \centering
        \includegraphics[width=9cm]{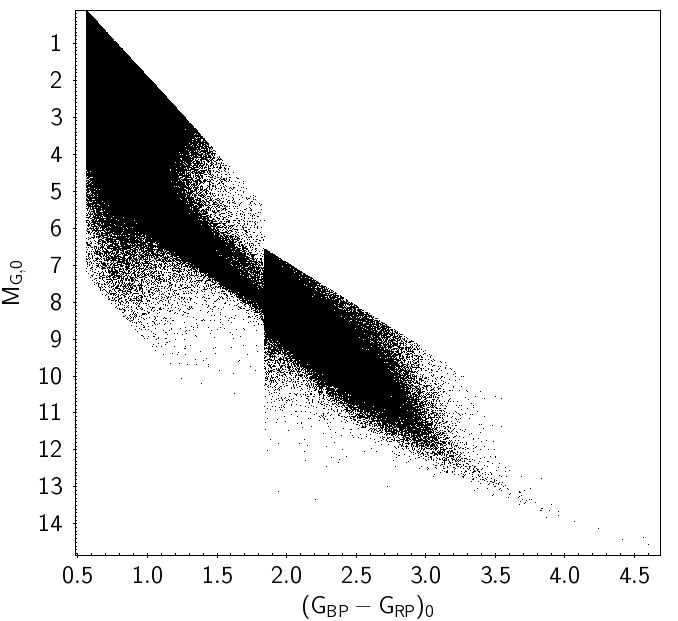}
        \caption{
    Combined selection of FGK and M stars in asPIC1.1.
        }
        \label{fig:asPIC1.1_selection}
\end{figure}

\begin{figure*}
        \centering
    \includegraphics[width=0.49\textwidth]{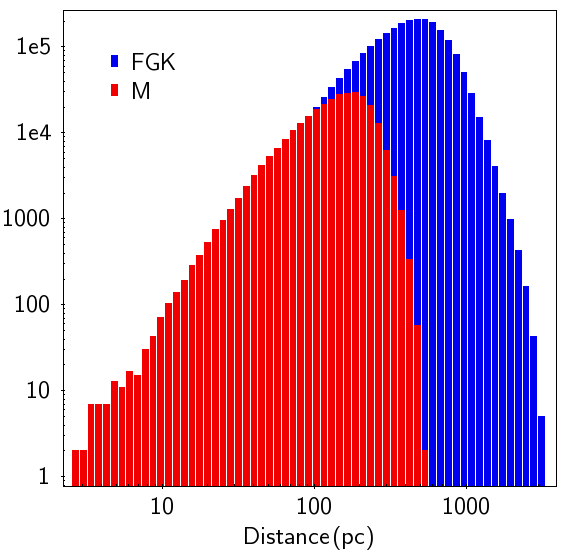}
    \includegraphics[width=0.49\textwidth]{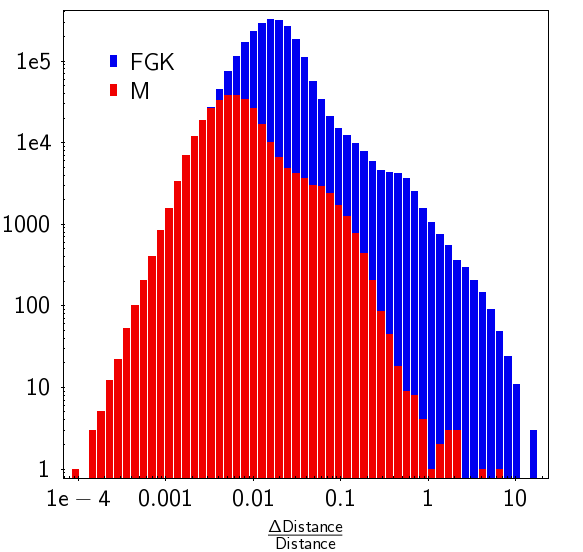}
        \caption{{\it Left:} Distributions of distances for M (red histogram) and FGK (blue histogram) stars as asPIC1.1. {\it Right:} Distribution of  relative errors on distances in asPIC1.1.
        }
        \label{fig:dist_param_P1P2P5}
\end{figure*}

\begin{figure*}
        \centering
        \includegraphics[width=17cm]{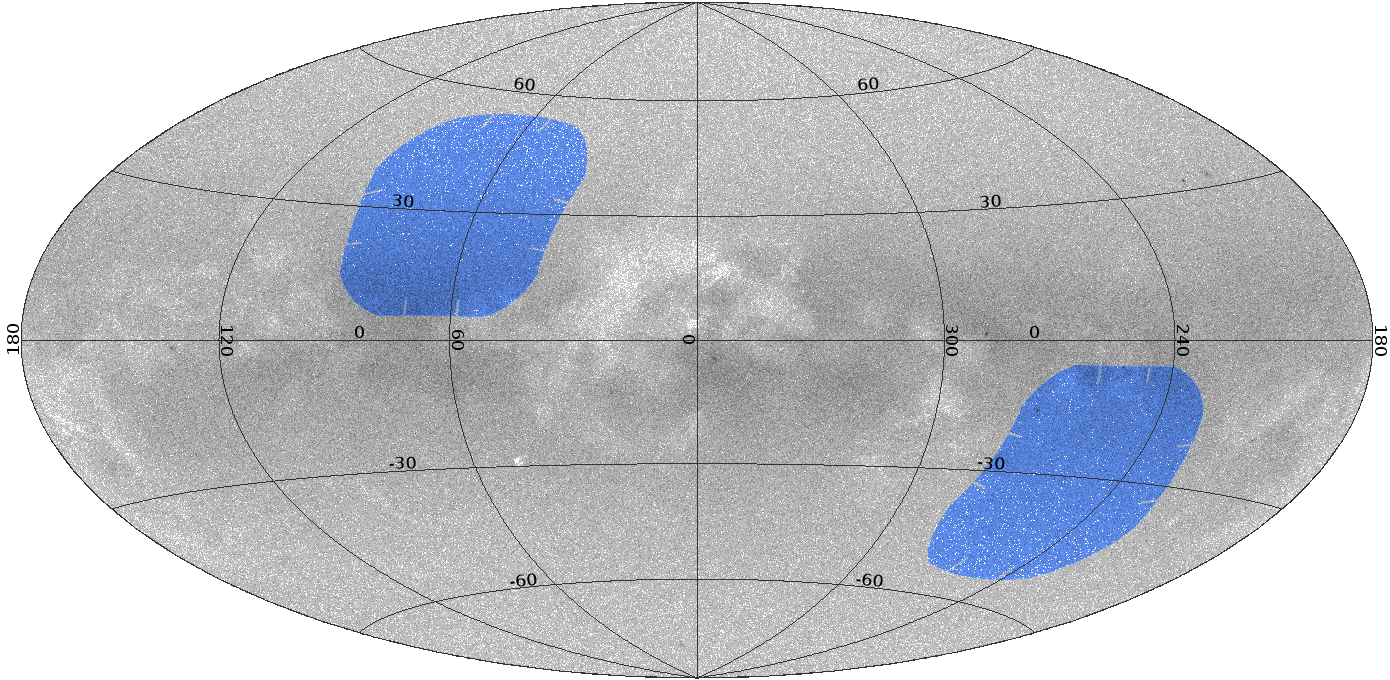}
        \caption{
    Distribution of asPIC1.1 dwarf and subgiant FGKM stars across the celestial sphere (grey points) in a Galactic coordinates reference system. The plot also shows the position of the two provisional long-duration fields (NPF and SPF). 
        }
        \label{fig:asPIC1.1}
\end{figure*}

Figure~\ref{fig:asPIC1.1_selection} illustrates the selection in the $M_{G,0}$ versus ({\it G$\rm_{BP}$-G$\rm_{BP}$})$_0$ diagram while Fig.~\ref{fig:dist_param_P1P2P5} (left panel) presents the distributions of distances for the M and FGK samples. The median values of M and FGK distances are 146\,pc and 428\,pc, respectively. The right panel of Fig.~\ref{fig:dist_param_P1P2P5} shows the distribution of relative distance errors for the different stellar samples. The error on the distance is calculated from the upper and lower distance limits reported by \citet{bailer2018} and is equal to the semi-difference of these values.
In terms of percentage errors in distance, the median values for the M and FGK stars are $0.6$ and $1.6$\%, respectively. Considering the median distances of the stellar samples reported 
above, this means that M and FGK distances are set with a precision of 0.9 pc and 6.8\,pc, respectively. 
Figure~\ref{fig:asPIC1.1} shows the distribution of the selected FGKM dwarf and subgiant stars across the celestial sphere in a Galactic coordinate reference system and the position of the two currently provisionally defined LOP fields
\citep{nascimbeni2016}.
These fields are 
centred at Galactic coordinates l=65$^{\circ}$, b=30$^{\circ}$
(RA=17$\rm^{h}$:40$\rm^{m}$:19.265$\rm^{s}$; DEC=+39$\rm^{\circ}$:35$\rm^{\prime}$:01.47$\rm^{\prime\prime}$) for the NPF and l=253$^{\circ}$, b=-30$^{\circ}$ for the SPF (RA=05$\rm^{h}$:47$\rm^{m}$:11.700$\rm^{s}$; DEC=-46$\rm^{\circ}$:23$\rm^{\prime}$:45.37$\rm^{\prime\prime}$).
The definitive choice of the LOP fields, to be frozen at the latest two years before launch, involves a complex optimisation task to merge several constraints and priorities of both engineering and scientific nature and will be discussed in Nascimbeni et al.~(in preparation). 
Nevertheless, the two provisional fields illustrated in Fig.~\ref{fig:asPIC1.1} already satisfy all PLATO requirements (Sect.~\ref{sec:stellar_samples}).

The final all-sky catalogue is named asPIC1.1 and contains 2 675 539 stars. There are 2 378 177 FGK dwarfs and subgiants ($V\leq$13) and 297 362 M-dwarfs ($V\leq$16).
In Appendix~\ref{sec:PIC1.0.0}, we describe an alternative selection of FGK dwarf and subgiant stars based on Johnson $B$ and $V$ photometry, which was used to construct the first version of this catalogue (asPIC1.0).
In Appendix~\ref{sec:catalogue}, we describe the catalogue content, while in Appendix~\ref{sec:implementation}, we present the details of the implementation procedure that has been adopted to construct asPIC1.1.

\section{Reddening and absorption}
\label{sec:reddening}

Because of the presence of interstellar matter, stars appear redder and fainter than they are. This has an impact on the selection of PLATO targets which is based on the use of photometric quantities like colours and magnitudes, as discussed in Sect.~\ref{sec:selection_criteria}. In particular, as the PLATO samples are magnitude limited, F-type stars, being intrinsically more luminous are located at larger distances on average than later type stars, and are therefore the most affected by reddening. By neglecting the
reddening and absorption correction we estimate that 100\% of the stars at the blue limit of our colour selection ({\it G$\rm_{RP}$-G$\rm_{BP}$}=0.56, Eq.~\ref{eq:P5}) would be contaminant dwarfs and subgiants (i.e. hotter than spectral type F5). It is therefore important to account for the impact of
interstellar matter
before selecting the target stars. 


Different approaches can be considered to determine reddening and extinction. One of the drivers for the choice of the best approach to follow for the PLATO input catalogue is to consider that the most important targets lie in the solar neighbourhood. Considering FGK targets, reddening should be estimated up to a distance of about 1\,kpc.

\begin{figure}
	\centering
    \includegraphics[width=4cm]{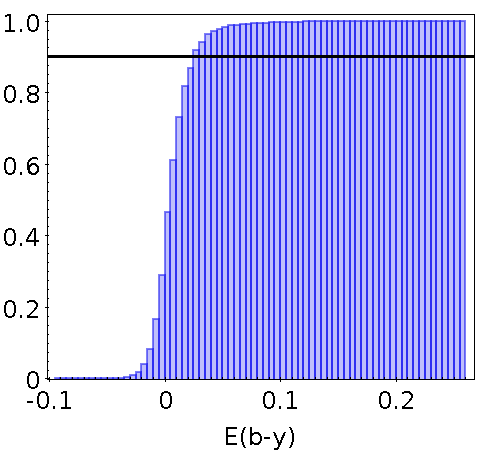}
    \includegraphics[width=4cm]{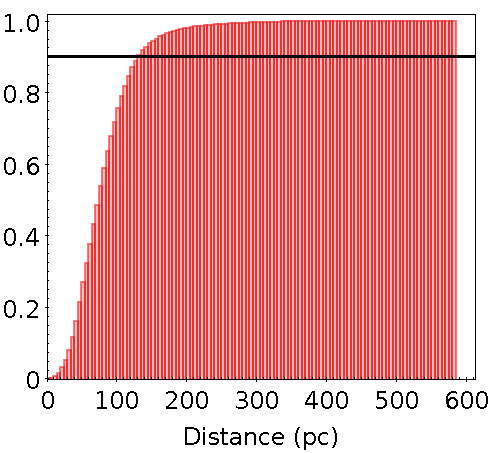}
	\caption{{\it Left:} cumulative distribution of reddening values in GCS.
	{{\it Right:} cumulative distribution of HIPPARCOS distances in GCS. Horizontal lines denotes the $90^\textrm{th}$ percentile of the cumulative distributions.}
	}
	\label{fig:gsc}
\end{figure}

\noindent
Figure~\ref{fig:gsc} shows the cumulative distributions of reddening (left) and HIPPARCOS distances (right) for all stars in the Geneva-Copenhagen Survey (GCS)\footnote{We used here the GCSII version \citep{holmberg2007} which reports the reddening.}. The figure shows that 90\% of these stars are within 150\,pc and the reddening is $E(b-y ) \leq 0.02
$ corresponding to $E(B-V) \leq 0.04$. In the GCS, reddening is estimated from the calibration by \citet{olsen1988} of the Str{\"o}mgren photometry intrinsic colour index $(b-y)$. The stated precision is 0.009\,mag. The same figure
also shows
that around $20\%$ of the stars have negative reddening, and are too close to
allow
a reliable reddening determination
At the small distances of PLATO targets, determining the reddening on a purely empirical photometric basis is challenging and requires measurements of  very high precision.
Such estimates can be obtained from measurement of interstellar neutral sodium absorption lines imprinted on spectra of background stars \citep[e.g.][]{Vergely2001}. Such measurements are best performed on early-type stars and extrapolation techniques can be used to reconstruct tomographic maps of the local interstellar medium in all directions. Such spectroscopic measurements have a superior precision than purely photometric estimates, but the empirical database is still limited \citep{welsh2010}, and the resulting reddening maps have a limited spatial extent ($<300$\,pc from the Sun) and would be largely incomplete for our samples.

\noindent
In general, spectroscopic measurements can be used to determine the effective temperature and therefore the intrinsic colours of stars, which can then be compared with the observed colours in several optical and near-infrared bands to infer colour excesses \citep{chen2019}. These methods suffer from the completeness problem mentioned above (although the availability of large spectroscopic surveys mitigates this problem),
from the
inhomogeneity of the spectroscopic samples adopted and
from
the different accuracies and precisions of the effective temperature determinations.

\begin{figure}
        \centering
    \includegraphics[width=8cm]{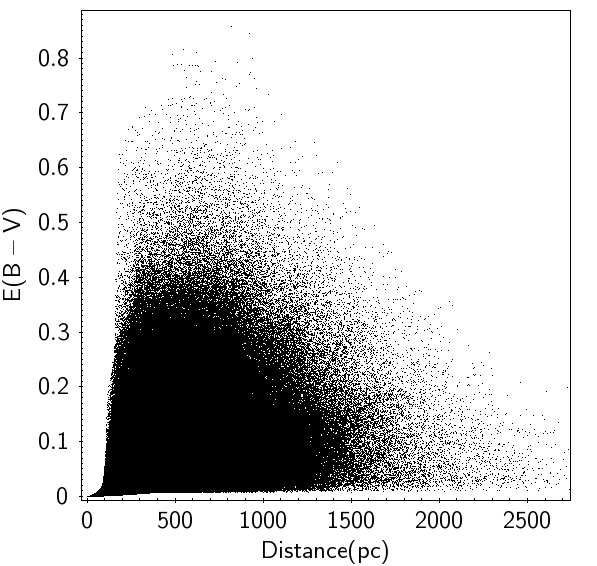}
        \caption{Reddening {\it E(B-V)} vs distance for all stars in asPIC1.1.
        }
        \label{fig:distebv}
\end{figure}

\noindent
A reasonable compromise among the different techniques can be obtained by coupling empirical determinations with appropriate priors derived from theoretical models or observational
constraints. Several authors began to produce 3D reddening maps using optical and/or near-infrared photometry and spectroscopy from large surveys adopting different assumptions on dust and stellar distributions \citep{green2018,gontcharov2010}. Among these, we chose the one recently presented in \citet{lallement2018}\footnote{Very recently \citet{lallement2019} published an updated version of the map, which is not included in this version of the catalogue but will be considered in the next version.} because of its accurate description of the local interstellar medium and its overall spatial coverage which
allowed us to determine the reddening for a large fraction of PIC stars in a homogeneous way. The 3D reddening map of \cite{lallement2018} is described in Appendix~\ref{sec:3d_reddening}.
The median reddening of asPIC1.1 stars inside the reddening map is equal to $E(B-V)$=0.04 and the median uncertainty is $\sigma E(B-V)$=0.02. For the M sample, $99.8\%$ of the stars are contained in the reddening map and, for the FGK sample, this figure is $81.8\%$. For the stars falling outside the map, reddening is calculated up to the edge of the map, and then a correction is added, as described in Sect.~\ref{sec:extension_reddening_map}.
Figure~\ref{fig:distebv} presents the {\it E(B-V)} versus distance diagram for all stars in asPIC1.1.

\section{Stellar parameters}
\label{sec:stellar_parameters}

This section describes the 
algorithms we used to estimate effective temperatures, radii, and masses.

\subsection{Intrinsic colour-effective temperature relation}
\label{sec:intrinsic_colour_effective_temperature}

We first determined the relationship between the intrinsic colour in the \textit{Gaia} bands and the effective temperature. For FGK spectral types, we used the stellar sample from \citet{casagrande2010}. Effective temperatures reported in this catalogue are calculated using the infrared flux method (IRFM).
We cross-matched this sample with \textit{Gaia} DR2 and considered only stars closer than 40\,pc from the Sun. \citet{casagrande2010} derived the IRFM effective temperatures for all stars with HIPPARCOS parallaxes and distances $\le 70$\,pc from the Sun assuming they are unaffected by reddening. To be consistent with this choice we assumed that the sample of calibration stars we used (all located within 40\,pc) have null reddening. The astrometric and photometric quality conditions expressed by Eq.~(1) and Eq.~(2) in \citet{arenou2018} were verified for all the selected stars\footnote{With the exception of one source, the bright spectroscopic binary HD112758, for which the astrometric condition expressed by Eq.~(1) of \citet{arenou2018} was not satisfied, but that was kept in the sample because it is known to host a low-stellar-mass companion
\citep{reffert2011}.}.
Our final list comprises 110 stars out of the 423 stars of \citet{casagrande2010}.\\
For M-dwarfs, we  considered the sample of \citet{mann2015}. All the 179 (see Sect.~\ref{sec:control_sample}) stars in the sample are located within 40 pc from the Sun. We also checked that Eq.~(1) and Eq.~(2) of \citet{arenou2018}
were satisfied, selecting in this way a total of 171 stars. Their effective temperatures were estimated from optical and near-infrared spectra.
Finally, for hot stars (T$_{\textrm{eff}}>$6510 K), we considered the temperature and colour estimates reported in \citet{pecaut2013}.

\begin{figure}
        \centering
        \includegraphics[width=\columnwidth]{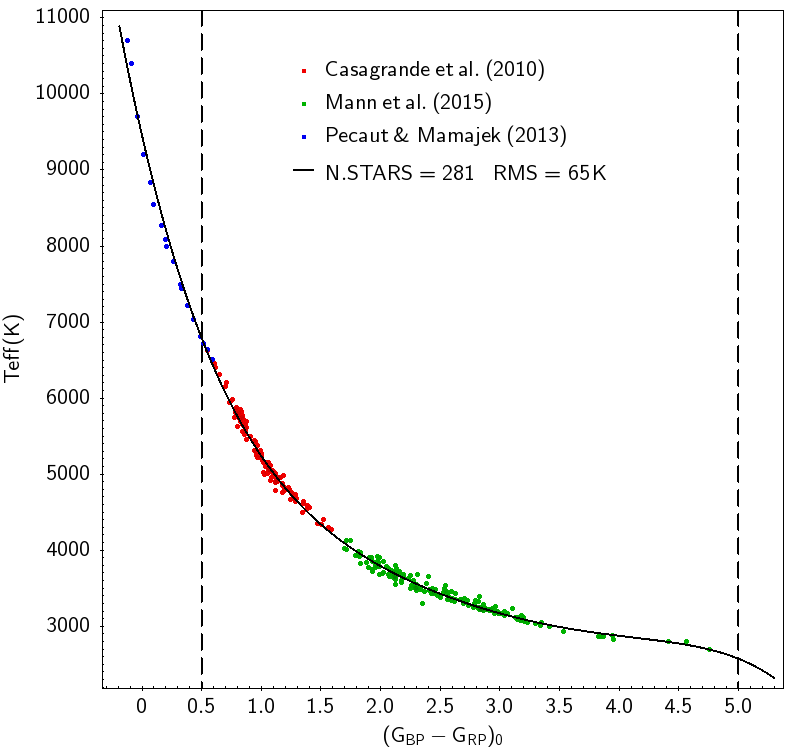}
        \caption{Relation between effective temperature T$_\textrm{eff}$ and intrinsic colour $\rm (G_\textrm{BP}-G_\textrm{RP})_0$. The continuous black line denotes our best-fit model (Eq.~ \ref{eq:Teff-1guess}), while the dashed vertical lines denote the limits of validity of the relation. Coloured dots represent the samples of \citet{casagrande2010}, \citet{mann2015}, and \citet{pecaut2013} discussed in the text in Sect.~{\ref{sec:intrinsic_colour_effective_temperature}}.
        }
        \label{fig:casagrande}
\end{figure}

As shown in Figure~\ref{fig:casagrande}, from the resulting sample of 281 stars we interpolated a single fifth-order polynomial relating the observed \textit{Gaia} colour of the calibration stars and their effective temperature. The coefficients of the fit are reported below, and the RMS of the residuals of the fit is equal to 65\,K

\begin{align}\label{eq:Teff-1guess}
T_\textrm{eff} (K) & = 9453.14 - 6859.40\,(G_\textrm{BP}-G_\textrm{RP})_0 + \nonumber \\ 
& + 3542.16\,(G_\textrm{BP}-G_\textrm{RP})_0^2 + \nonumber \\
& - 1053.09\,(G_\textrm{BP}-G_\textrm{RP})_0^3+165.635\,(G_\textrm{BP}-G_\textrm{RP})_0^4 + \nonumber \\
& -10.5672\,(G_\textrm{BP}-G_\textrm{RP})_0^5, 
\end{align}

\noindent
where the relation is valid for $0.5<$(G$_\textrm{BP}-G_\textrm{RP})<5$.
Hereafter, we use Eq.~\eqref{eq:Teff-1guess} as the relation between the intrinsic colour and the effective temperature.
Using Eq.~(\ref{eq:Teff-1guess}) we obtained a first guess of the effective temperature of all asPIC stars. These stars are generally located further, at larger distances than the sample of calibration stars, and therefore reddening cannot be neglected. 

As reported in Section~\ref{sec:reddening}, we estimate reddening using the reddening map of \citet{lallement2018}. The customised algorithm described in Section.~\ref{sec:3d_reddening} furnishes the interpolated reddening $E(B-V)$ for any 3D star position. As input, it requires the
Galactic longitude $l$ and the Galactic latitude $b$ which are taken directly from the input catalogue along with the distance from the \citet{bailer2018} catalogue.

\subsection{Conversion from \texorpdfstring{$E(B-V)$}{} to \texorpdfstring{$E(G_\textrm{BP}-G_\textrm{RP})$}{} and determination of extinction \texorpdfstring{$A_\textrm{G}$}{}}
\label{sec:conversion}

The $E(B-V)$ reddening is converted to $E(G_\textrm{BP}-G_\textrm{RP})$ and to the extinction $A_\textrm{G}$ in the \textit{Gaia} $G$-band as follows:

\begin{align}
    & A_\textrm{G}=R_\textrm{G}\,E(B-V),\\
    & A_\textrm{BP}=R_\textrm{BP}\,E(B-V),\\
    & A_\textrm{RP}=R_\textrm{RP}\,E(B-V),
\end{align}

\noindent
where the \textit{Gaia} extinction coefficients
depend on the effective temperature
as follows

\begin{align}
R_\textrm{G}  = & -0.5335 + 12.9373\,(T_4) - 13.9514\,(T_4)^2 + \nonumber \\
& - 13.8012\,(T_4)^3 + 40.9902\,(T_4)^4 - 23.6648\,(T_4)^5,  \label{eq:rrg}\\
R_\textrm{BP} = & -2.4689 + 59.5802\,(T_4) - 253.9922\,(T_4)^2 + \nonumber \\
& + 526.5333\,(T_4)^3 - 523.9970\,(T_4)^4 + \nonumber \\
& + 201.2829\,(T_4)^5, \label{eq:rbp}\\
R_\textrm{RP} = & 0.0407 + 9.8825\,(T_4) - 16.8207\,(T_4)^2 + \nonumber\\
& + 9.1283\,(T_4)^3 + 2.5051\,(T_4)^4 - 2.4158\,(T_4)^5,\label{eq:rrp}\\
\nonumber
\end{align}

\noindent
and

\begin{equation}
    T_4 = 10^{-4}\,(T_\textrm{eff}).
\end{equation}

\begin{figure*}
        \centering
        \includegraphics[width=0.32\textwidth]{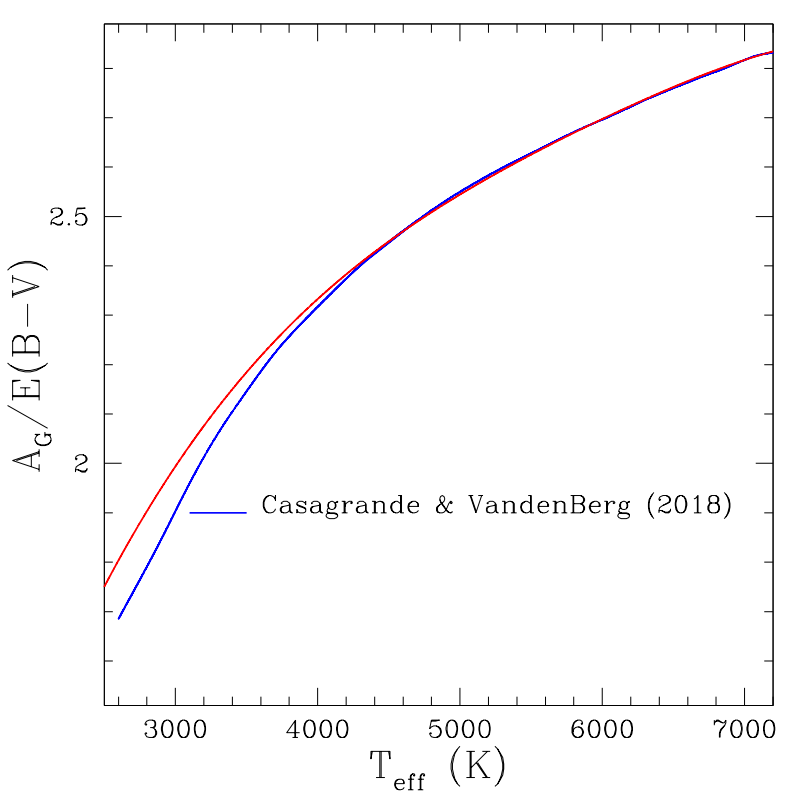}
    \includegraphics[width=0.32\textwidth]{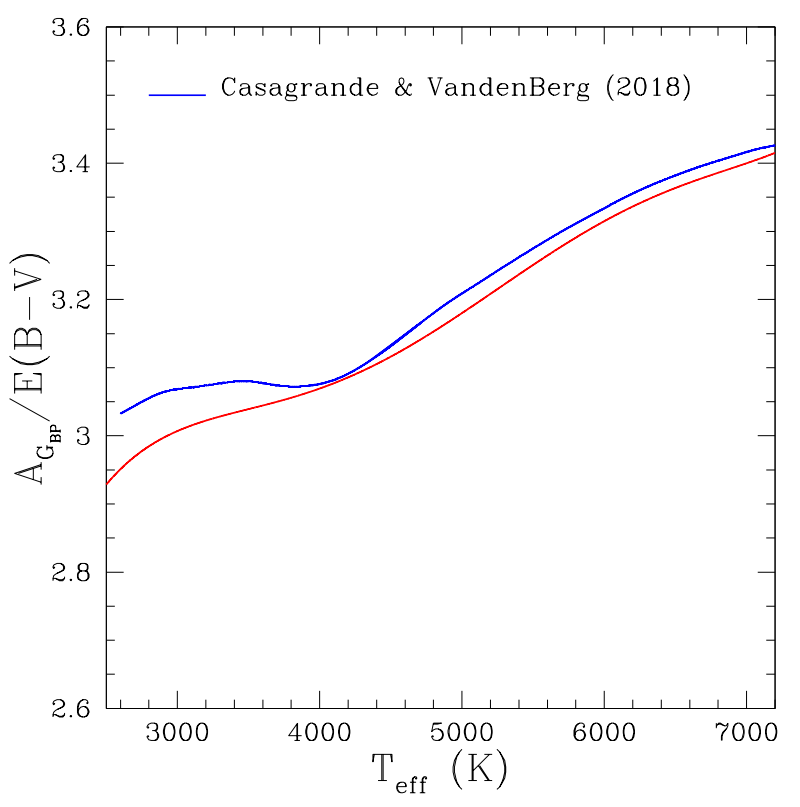}
    \includegraphics[width=0.32\textwidth]{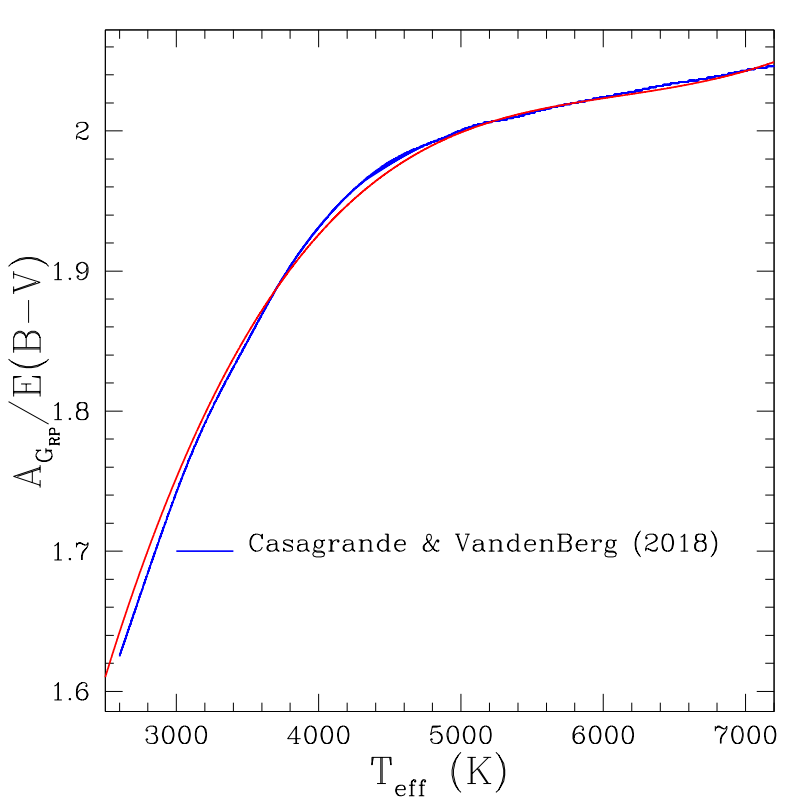}
        \caption{Relations between $A_G/E(B-V)$ (left), $A_{G_{BP}}/E(B-V)$ (middle), $A_{G_{RP}}/E(B-V)$ (right) and effective temperature T$_{\rm eff}$ derived with the procedure described in the text (Sect.~\ref{sec:conversion}). The red lines represent our interpolating relations and the blue lines models from \cite{casagrande2018}.}
        \label{fig:extcoeff}
\end{figure*}

\noindent
Such relations were derived using Padova stellar models \citep{bressan2012}, assuming
solar metallicity, solar composition, and temperatures within the interval $2500\,\textrm{K} <T_\textrm{eff} < 7200\,\textrm{K}$. In particular, 
we modelled the ratio of the absorption in a given photometric band to the reddening $E(B-V)$ considering an unreddened 1 Gyr isochrone, 
and the same isochrone reddened using $A_V$=0.5 mag, where reddening was applied on a star-by-star basis. We then interpolated apparent magnitudes versus effective temperature on the same temperature grid for both the reddened and unreddened isochrones, and calculated the absorption on each \textit{Gaia} band as a function of temperature. 
In Figure~\ref{fig:extcoeff}, we show our interpolating relations presented in Eqs.~(\ref{eq:rrg}), (\ref{eq:rbp}), and (\ref{eq:rrp})
along with the results obtained using the bolometric correction program of \citet{casagrande2018}\footnote{https://github.com/casaluca/bolometric-corrections}. 
In particular, we used this program to calculate the bolometric corrections for unreddened and reddened stars interpolating across the grid of log g and effective temperatures of our adopted isochrone (assuming solar metallicity and solar composition), and from the bolometric corrections we calculated the ratios between the extinction in each band and the reddening {\it E(B-V)} as a function of the effective temperature 
 \footnote{ $\frac{A\rm_{\gamma}}{E(B-V)}$=
 $\frac{(BC_{A\textrm{v}=0}-BC_{A\textrm{v}})}{A\textrm{v}}\times$3.1, where $\gamma=G\rm_{BP}$,$G$,$G\rm_{Rp}$ and Av=0.5.}. The results we obtained from our procedure agree well with those of \citet{casagrande2018} in general. Some small differences are visible, particularly for late-type stars and for the {\it G} and {\it G$_{BP}$} bands which may be explained by differences in the underlying model atmospheres, because the Padova group colours and magnitudes are based on the libraries of stellar spectra of \cite{castelli2003}, whereas \citet{casagrande2018} used the MARCS library of theoretical stellar fluxes \citep{gustafsson2008}. 
  
\subsection{Correction for reddening and extinction}
\label{sec:correct}

The observed $G_\textrm{BP}-G_\textrm{RP}$ colour and $G$ magnitude were corrected for reddening and extinction to obtain the intrinsic colour $(G_\textrm{BP}-G_\textrm{RP})_0$ and the intrinsic magnitude $G_0:$

\begin{align}
    & (G_\textrm{BP}-G_\textrm{RP})_0 = (G_\textrm{BP}-G_\textrm{RP})-E(G_\textrm{BP}-G_\textrm{RP}),\label{eq:intrinsic_colour}\\
    &\nonumber \\
    &G_0 = G-A_\textrm{G}.
\end{align}

\subsection{Effective temperature: the iteration cycle}

We recalculated the effective temperature using Eq.~(\ref{eq:Teff-1guess}) and the new intrinsic colour $(G_\textrm{BP}-G_\textrm{RP})_0$ in Eq.~(\ref{eq:intrinsic_colour}).
We then used the new effective temperature to estimate the reddening
coefficients (Section~\ref{sec:conversion}) and then a new estimate of the intrinsic colour (Section~\ref{sec:correct}).
We then used this to estimate  the effective temperature again.
A subsequent iteration lead to effective temperatures that differ by less than 10\,K from last estimate; therefore two iterations were sufficient.

\subsection{Bolometric correction in the \texorpdfstring{$G-$}{}band, \texorpdfstring{$BC_\textrm{G}$}{}}

The bolometric correction of the $G$-band is obtained from the relations given in \cite{andrae2018}:

\begin{align}
    BC_\textrm{G} & = 6.0 \times 10^{-2} + 6.731 \times 10^{-5}\,\Delta T_\textrm{eff} -6.647 \times 10^{-08}\,\Delta T_\textrm{eff}^2 + \nonumber \\
    & + 2.859 \times 10^{-11}\,\Delta T_\textrm{eff}^3 - 7.197 \times 10^{-15}\,\Delta T_\textrm{eff}^4,
\end{align}

\noindent
for a temperature interval 4000 K $\leq T_\textrm{eff} \leq 8000$ K and

\begin{align}
    BC_\textrm{G} & =1.749 + 1.977 \times 10^{-3} \,\Delta T_\textrm{eff}+ 3.737\times 10^{-7}\,\Delta T_\textrm{eff}^2 + \nonumber \\
    & - 8.966\times 10^{-11}\,\Delta T_\textrm{eff}^3 -4.183\times 10^{-14}\,\Delta T_\textrm{eff}^4,
\end{align}

\noindent
for a temperature interval 3300 K $\leq T_\textrm{eff} < 4000$ K and where $\Delta T_\textrm{eff}~=~T_\textrm{eff}~-~T_{\textrm{eff},\odot}$ and $T_{\textrm{eff},\odot} =5772$\,K.

\subsection{Determination of the absolute magnitude \texorpdfstring{$M_\textrm{G,0}$}{} and of the absolute luminosity \texorpdfstring{$L$}{}}

The intrinsic absolute magnitude $M_\textrm{G,0}$ and the luminosity $L$ were calculated as follows:

\begin{equation}
    M_\textrm{G,0} = G_0 - 5\,\log_{10}\,d + 5,
\end{equation}

\begin{equation}
    \frac{L}{L_\odot}=10^{-0.4\,(M_\textrm{G,0} + BC_\textrm{G} - M_{\textrm{BOL}_{\odot}})},
\end{equation}

\noindent
where $d$ is the distance of the star in parsecs from \citet{bailer2018} and $M_{\textrm{BOL}_{\odot}}$ = 4.74 and $L_\odot =3.828 \times 10^{26}$\,W.

\subsection{Determination of the stellar radius}

We then calculated  the stellar radius R  from the Stefan-Boltzmann law:

\begin{equation}
    \frac{R}{R_\odot}=\left ( \frac{T_\textrm{eff}}{T_{\textrm{eff},\odot}} \right )^{-2}\, \sqrt{\frac{L}{L_\odot}}.
    \label{eq:stellar_radius}
\end{equation}

\noindent
The presence of 
absorption and reddening 
influences 
the determination of the stellar radius because 
it affects both 
the estimation of the effective temperature and of the absolute luminosity.
Therefore it is important to apply dereddening procedures like those
we describe above. 
Reddening and absorption have two opposite and partially compensating effects on the stellar radius. A more detailed discussion about this point is provided in Appendix~\ref{sec:impact_of_reddening_on_stellar_radius}.

\subsection{Determination of the stellar mass}

The stellar mass was determined using the following relation from \cite{moya2018}:

\begin{equation}
    \log \left ( \frac{M}{M_\odot} \right ) = -0.119 + 2.14 \times 10^{-5}\, T_\textrm{eff} + 0.1837 \times \log \left ( {\frac{L}{L_\odot}} \right ).
\end{equation}

\noindent
The range of validity is $4780\, \textrm{K} \le T_\textrm{eff} \le 10990\, \textrm{K}$ and $-0.717 \le \log \left ( L / L_\odot \right )  \le 2.01$. This relationship is valid for dwarfs and subgiant stars.

\begin{figure*}
        \centering
    \includegraphics[width=0.32\textwidth]{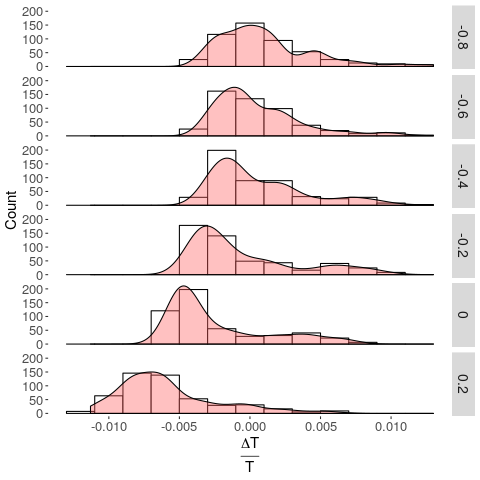}
    \includegraphics[width=0.32\textwidth]{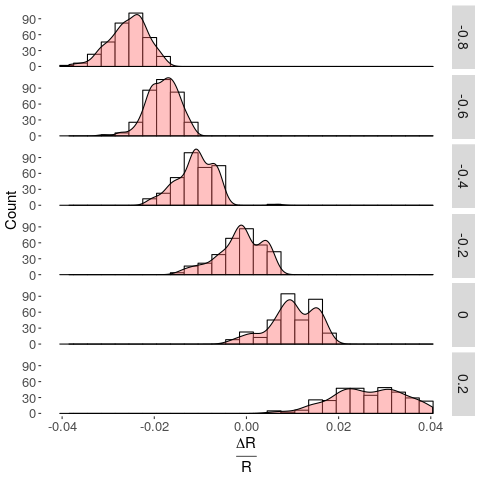}
  \includegraphics[width=0.32\textwidth]{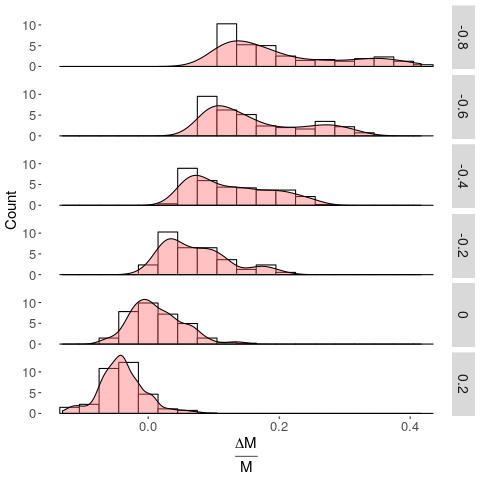}
        \caption{Relative differences between stellar parameters estimated using  asPIC1.1 and theoretical parameters from stellar isochrones in different metallicity intervals. The rightmost labels of each panel denote the centres of the [M/H] metallicity intervals having a width of $\pm$0.1 dex as also listed in Table~\ref{tab:reldiff_met}.}
        \label{fig:reldiff_met}
\end{figure*}

\subsection{Impact of metallicity on the determination of stellar  parameters}
\label{sec:metal_param}

The procedure used to determine stellar parameters described in Sect.~\ref{sec:stellar_parameters} does not take into account metallicity because we do not have
metallicity measurements for most of our stars.
In this section, we estimate the impact that metallicity has on the  determination of stellar parameters. We retrieved  a grid of isochrones with age between 7$<$log(Age/yr)$<$10 in steps of 0.2 dex and metallicity between -1<[M/H]<0.3 in steps of 0.1 dex from the Padova stellar database. For each model in the grid, we determined the stellar parameters with the procedure described in Sect.~\ref{sec:stellar_parameters} and compared them with the theoretical parameters reported in the isochrones for six different intervals of metallicity. \ In particular, we calculated the relative differences ($\rm \frac{\Delta T}{T}$,
$\rm \frac{\Delta R}{R}$, $\rm \frac{\Delta M}{M}$) between our parameters (T$\rm_{eff, asPIC1.1}$, Radius$\rm_{asPIC1.1}$, Mass$\rm_{asPIC1.1}$) and the theoretical parameters (T$\rm_{eff, ISO}$, Radius$\rm_{ISO}$, Mass$\rm_{ISO}$)

\begin{equation}
\rm \frac{\Delta T}{T}=\frac{T_{eff, asPIC1.1}-T_{eff, ISO}}{T_{eff, ISO}},    
\end{equation}

\begin{equation}
\rm \frac{\Delta R}{R}=\frac{Radius_{asPIC1.1}-Radius_{ISO}}{Radius_{ISO}},
\end{equation}

\begin{equation}
\rm \frac{\Delta M}{M}=\frac{Mass_{asPIC1.1}-Mass_{ISO}}{Mass_{ISO}}.
\end{equation}

\begin{table}
        \centering
   \caption{
    Median relative differences (as a  percentage) and standard deviations between asPIC1.1 estimated stellar parameters and theoretical stellar parameters in different metallicity intervals.
    }   
        \begin{tabular}{cccc}
        \hline
     [M/H] & $\rm\frac{\Delta T}{T}$(\%) & $\rm\frac{\Delta R}{R}$(\%) & $\rm\frac{\Delta M}{M}$(\%)\\
    \hline
     -0.9$\leq$[M/H]$<$-0.7 & 0.02$\pm$0.33 & -2.5$\pm$0.4 & 17$\pm$9 \\
     -0.7$\leq$[M/H]$<$-0.5 & -0.04$\pm$0.30 & -1.8$\pm$0.3 &  14$\pm$7\\
     -0.5$\leq$[M/H]$<$-0.3 & -0.06$\pm$0.33 & -1.1$\pm$0.4 &  11$\pm$6\\
     -0.3$\leq$[M/H]$<$-0.1 & -0.21$\pm$0.38 & -0.1$\pm$0.5 &  6$\pm$5\\
     -0.1$\leq$[M/H]$<$0.1 & -0.41$\pm$0.36 & 1.0$\pm$0.5 &  0.5$\pm$3.9\\
      0.1$\leq$[M/H]$\leq$0.3 & -0.66$\pm$0.36 & 2.6$\pm$0.8 &  -4.1$\pm$3.5\\
    \hline
    \end{tabular}
    \label{tab:reldiff_met}
\end{table}

\noindent
The results are presented in Fig.~\ref{fig:reldiff_met} 
and in Table~\ref{tab:reldiff_met}. We find that the effective temperature has a weak dependence on metallicity with relative differences ranging from -0.6$\%$ to
0.02$\%$. The radius presents relative differences of between -2.5$\%$ and 2.5$\%$ while the mass shows the largest relative differences ranging from -4$\%$ to 17$\%$. We note that the mass is also the most uncertain among the theoretical parameters associated to the isochrones. Both the temperature and the mass relative differences are negatively correlated with the metallicity while the radius is positively correlated. We also calculated the standard deviations of the relative differences in each metallicity interval. These differences range between 0.3$\%$ and 0.4$\%$ for the temperature, between 0.3$\%$ and 0.8$\%$ for the radius, and between 4$\%$ and 9$\%$ for the mass.

\subsection{Errors estimate}
\label{sec:errors}

\begin{figure*}
        \centering
    \includegraphics[width=0.32\textwidth]{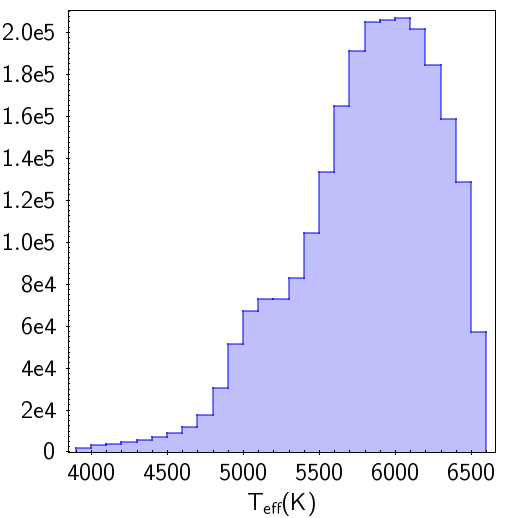}
    \includegraphics[width=0.32\textwidth]{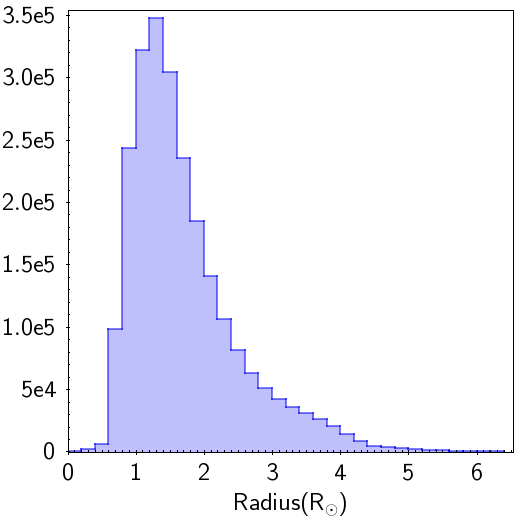}
  \includegraphics[width=0.32\textwidth]{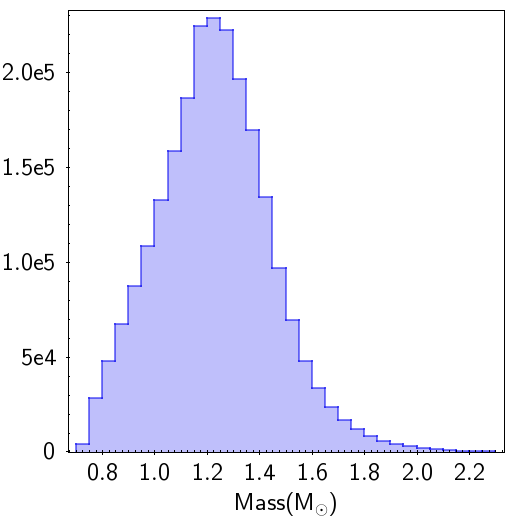}
        \caption{Distributions of effective temperature (left panel), radius (middle panel), and mass (right panel) for the FGK sample.}
        \label{fig:dist_param_P5}
\end{figure*}

We determined the errors on stellar parameters  using Monte Carlo simulations, perturbing all observing quantities according to their associated errors, and assuming a Gaussian perturbation.
Beyond colours, magnitudes, distances, and reddening, we also  perturbed the bolometric corrections adopting the errors reported by \citet{andrae2018}. The error on the reddening was taken from the reddening map of \citet{lallement2018} for stars falling inside its boundaries. For stars falling outside this map, the error on the reddening was assumed equal to twice the 
error on the median reddening of all stars inside the map, which is equal to $\sigma\,E(B-V)$=0.04  
(see also Sect.~\ref{sec:3d_reddening}).
The effective temperature deduced by the colour--effective temperature relation was perturbed considering a conservative error of 200\,K on its determination.
We performed 100 simulations for each star and then calculated the errors on the effective temperature, mass, and radius as the half interval between the 16$\rm^{th}$ and the 84$\rm^{th}$ percentile of the cumulative distribution of the simulated values.
The median value of the effective temperature error is  227 K (4$\%$), and this value is $0.1$ R$_{\odot}$ (7$\%$) for the radius, and $0.1$ M$_{\odot}$ (8$\%$) for the mass.

\subsection{Stellar parameter distributions}

In Fig.~\ref{fig:dist_param_P5} we present the distributions of effective temperature, radius, and mass for the FGK sample. 
We note that stellar parameters have not yet been calculated for the M dwarfs sample. They will be included in the next version of the catalogue.\\

\begin{table*}
        \centering
        \footnotesize
        \begin{tabular}{cccccc}
    \hline
     Catalogue & $\Delta T_\textrm{eff}$ & $\rm\Delta R_\star$ & $\rm\Delta M_\star$ & $\rm\Delta log\,g$ & $\Delta E(B-V)$ \\
             & $\rm(K)$ & $\rm(R_\odot)$ & $\rm(M_\odot)$ & (dex) &  (mag) \\
     \hline
    TIC (CTL$\rm_{v8}$) & $-100 \pm 300$ & $0.05 \pm 0.07$ & $0.05 \pm 0.17$ & - & $-0.002 \pm 0.055$ \\
    GALAH DR2 & 4$\pm$100 & - & - & 0.1$\pm$0.2 & - \\
    HARPS & -70$\pm$90 & - & - & -0.03$\pm$0.15 & - \\
    APOKASC (SDSS) & -130$\pm$90 & 0.1$\pm$0.2 & 0.1$\pm$0.1 & -0.02$\pm$0.06 & - \\
    APOKASC (ASPCAP) & 90$\pm$130 & 0.1$\pm$0.1 & 0.1$\pm$0.1 & -0.01$\pm$0.06 & - \\
    \hline
        \end{tabular}
    \caption{Median differences and standard deviation of the differences between asPIC1.1 and the stellar parameters and interstellar reddening values from other catalogues.}
    \label{tab:differences}
\end{table*}

\begin{figure*}
        \centering
    \includegraphics[width=0.49\textwidth]{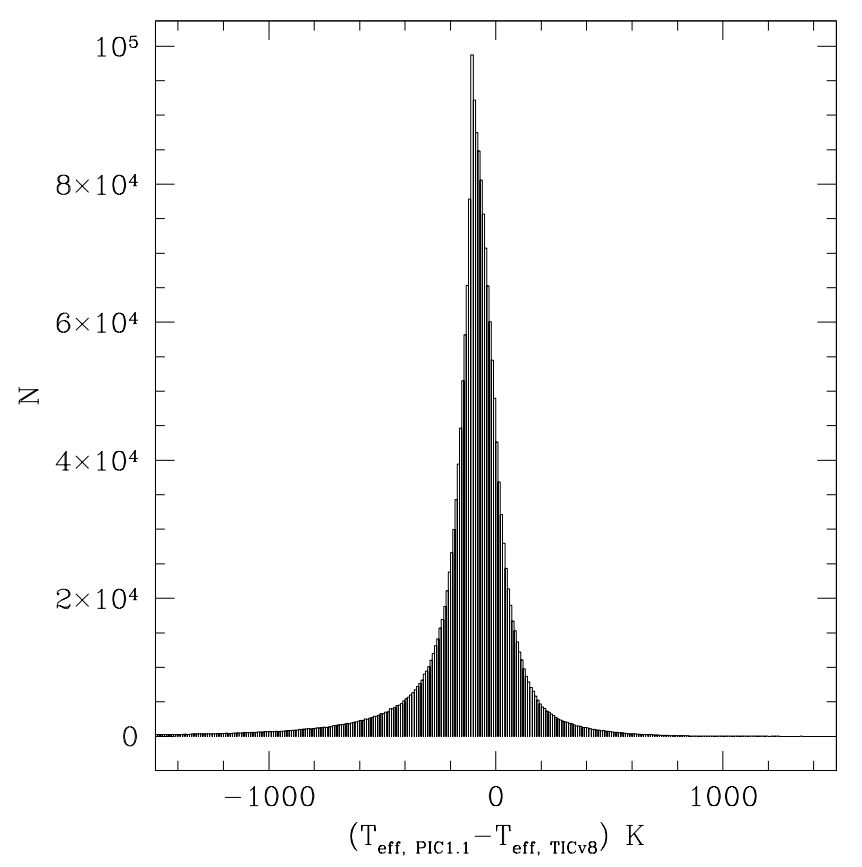}
    \includegraphics[width=0.49\textwidth]{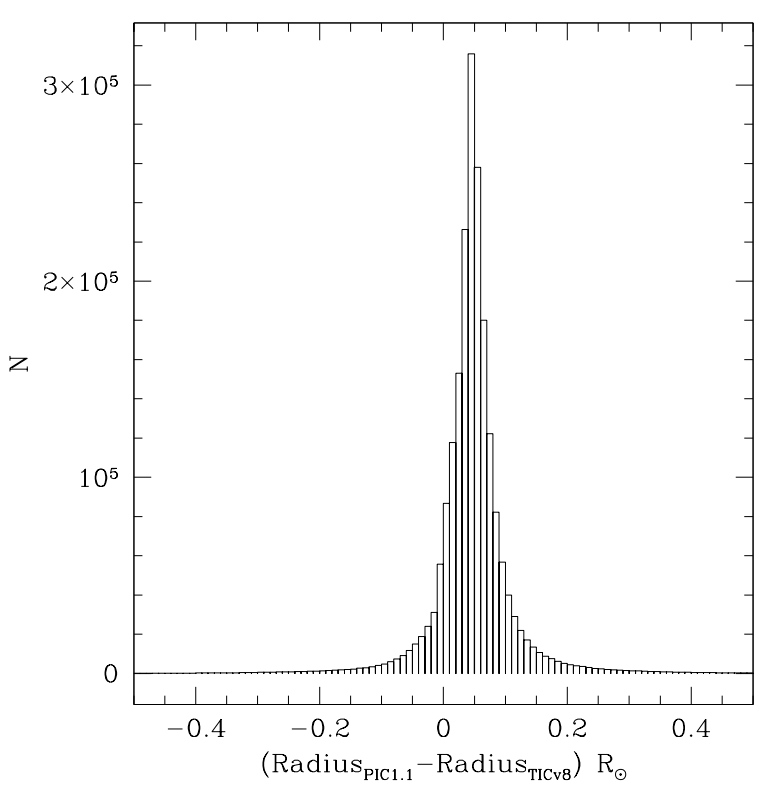}
    \includegraphics[width=0.49\textwidth]{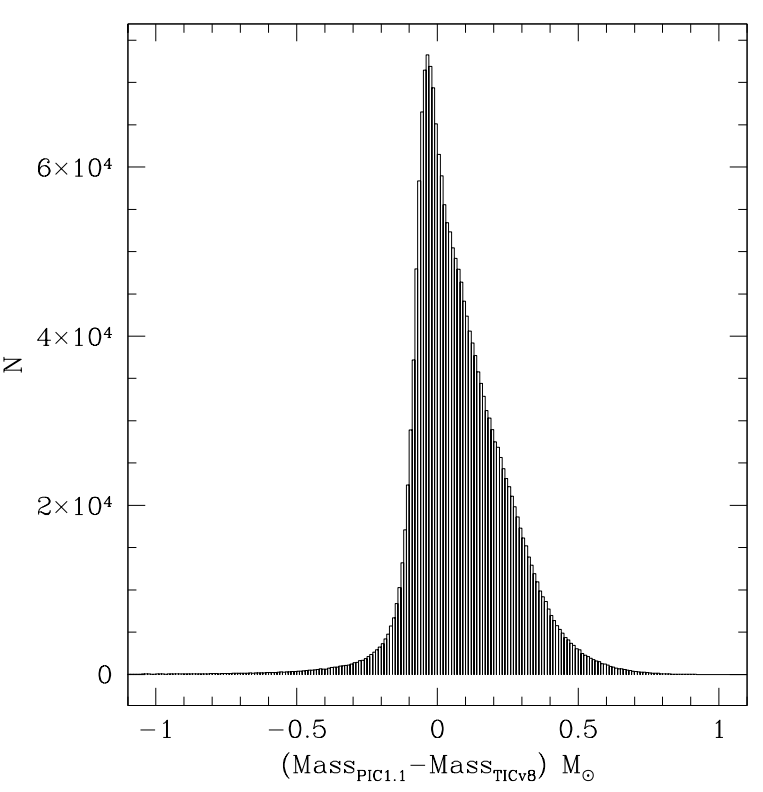}
    \includegraphics[width=0.49\textwidth]{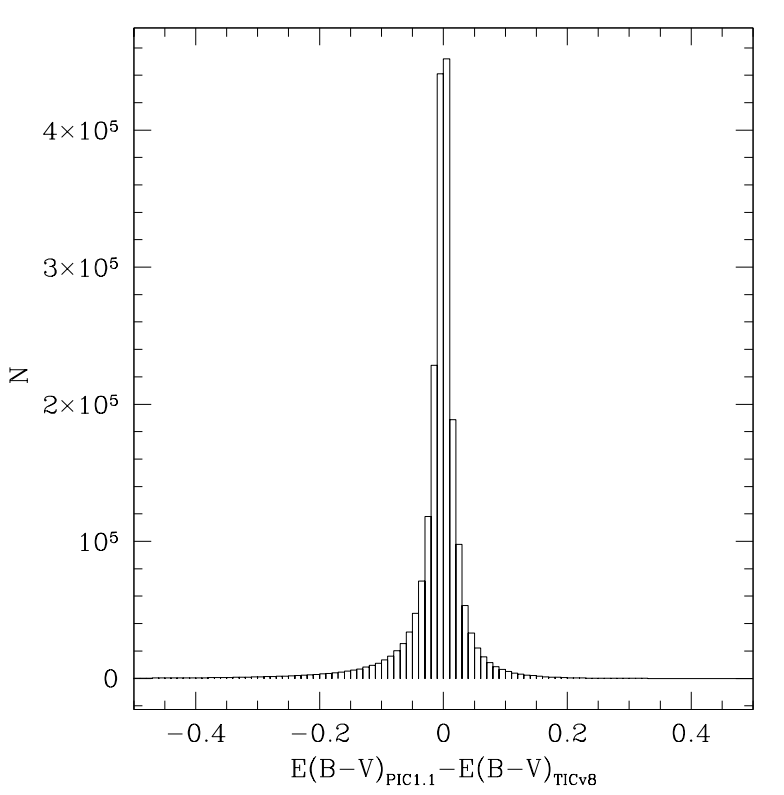}
        \caption{Distributions of effective temperature, stellar radius, mass, and interstellar reddening differences between asPIC1.1 and CTL.
        }
        \label{fig:comparison_TIC}
\end{figure*}

\begin{figure*}
        \centering
    \includegraphics[width=0.49\textwidth]{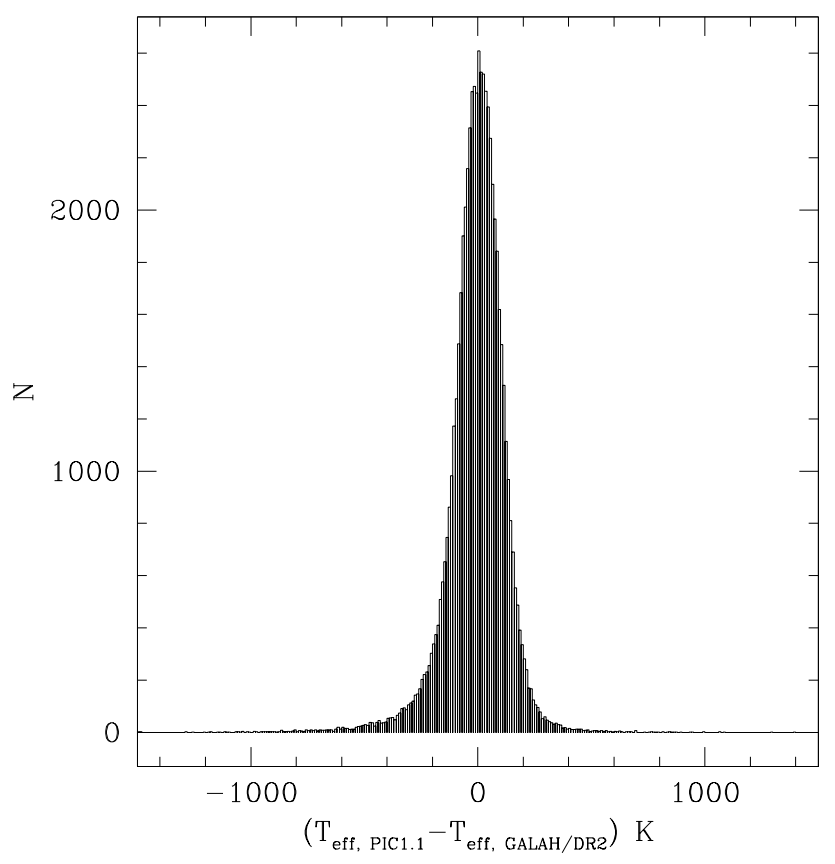}
    \includegraphics[width=0.49\textwidth]{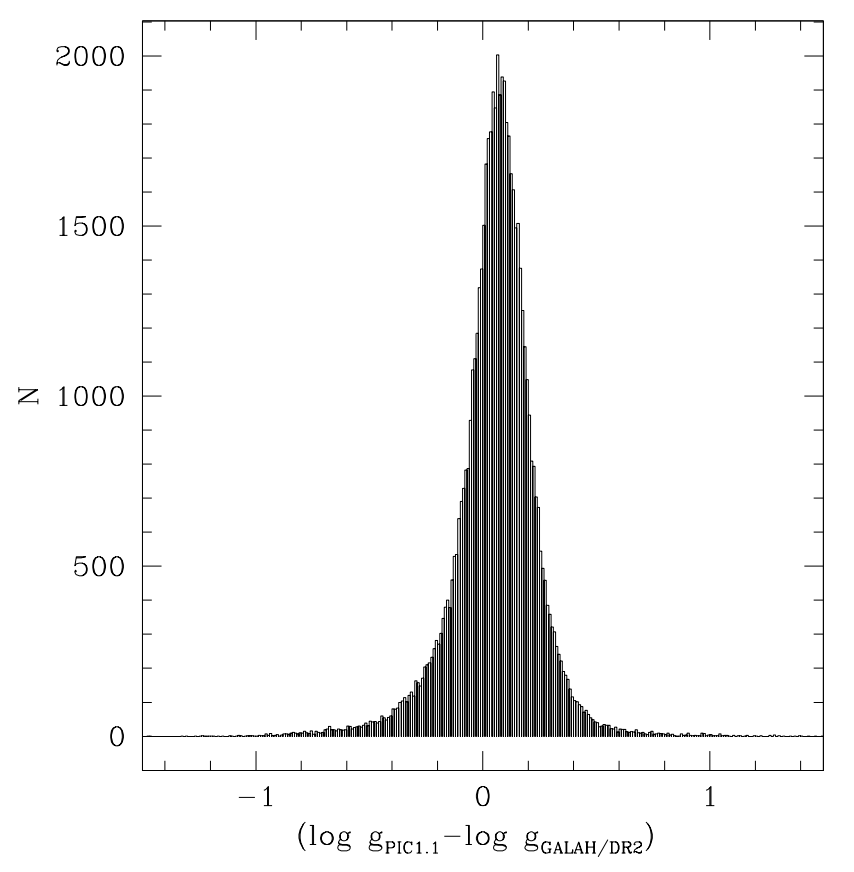}
        \caption{Distributions of effective temperatures, and stellar gravity differences between asPIC1.1 and GALAH/DR2.
        }
        \label{fig:comparison_GALAHDR2}
\end{figure*}

\begin{figure*}
        \centering
    \includegraphics[width=0.49\textwidth]{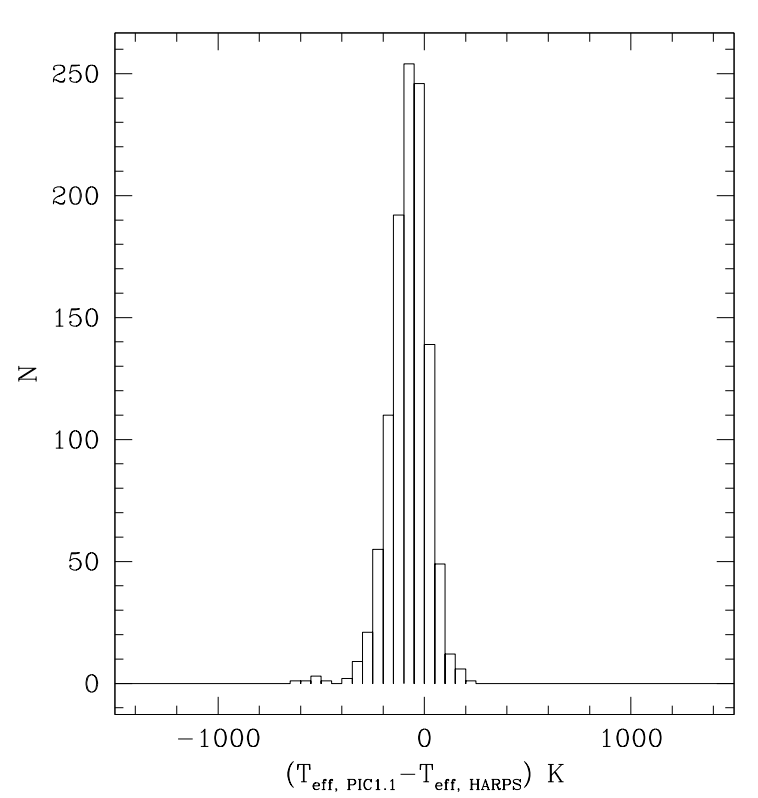}
    \includegraphics[width=0.49\textwidth]{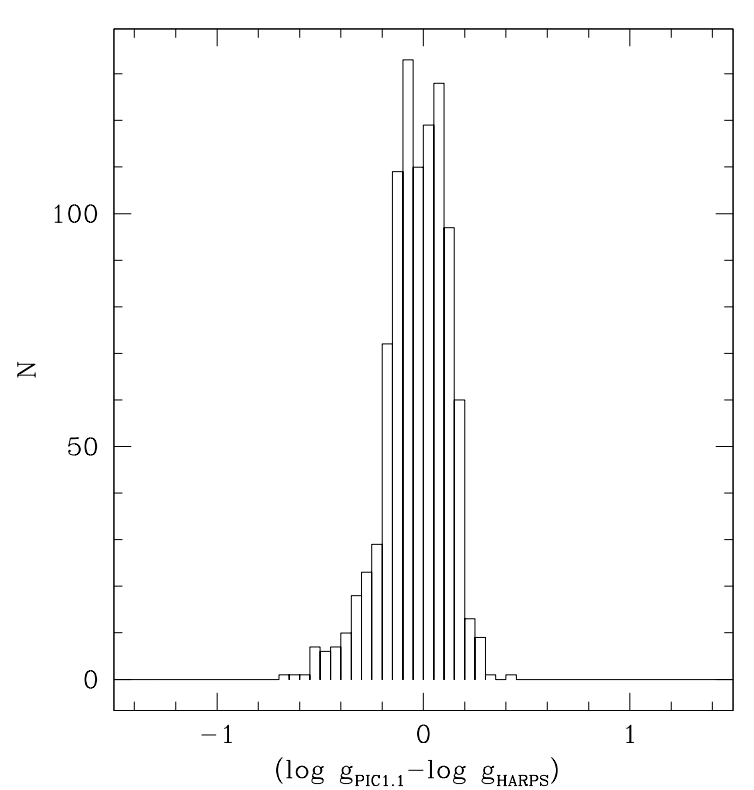}
        \caption{Distributions of effective temperatures, and stellar gravity differences between asPIC1.1 and the HARPS sample.}
        \label{fig:comparison_HARPS}
\end{figure*}

\begin{figure*}
        \centering
    \includegraphics[width=0.49\textwidth]{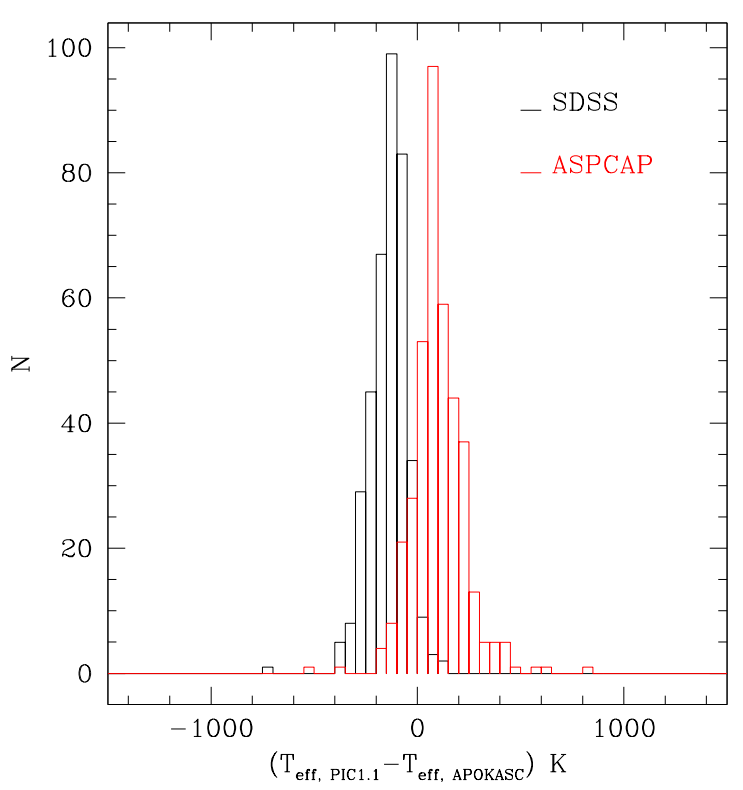}
    \includegraphics[width=0.49\textwidth]{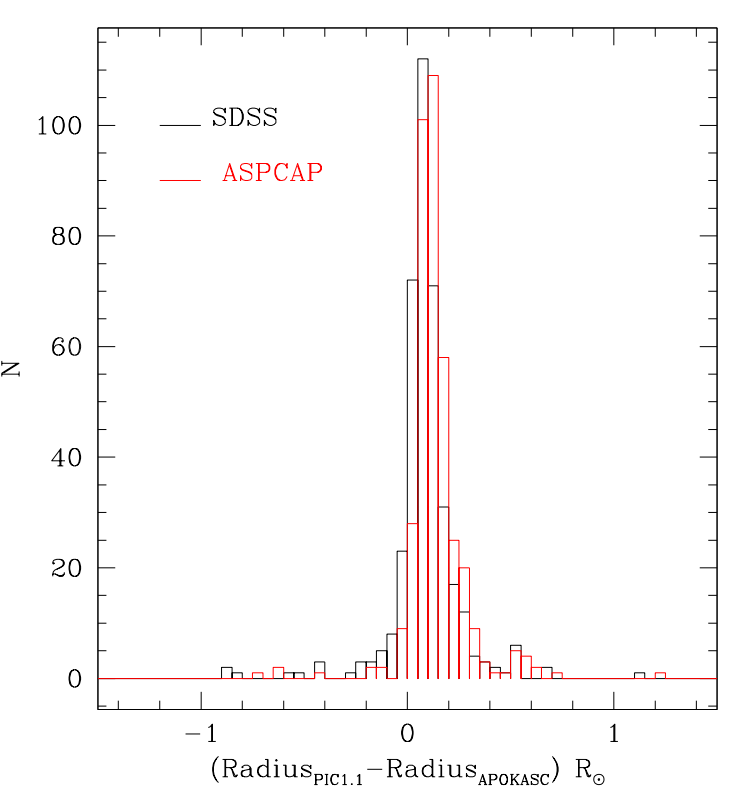}
    \includegraphics[width=0.49\textwidth]{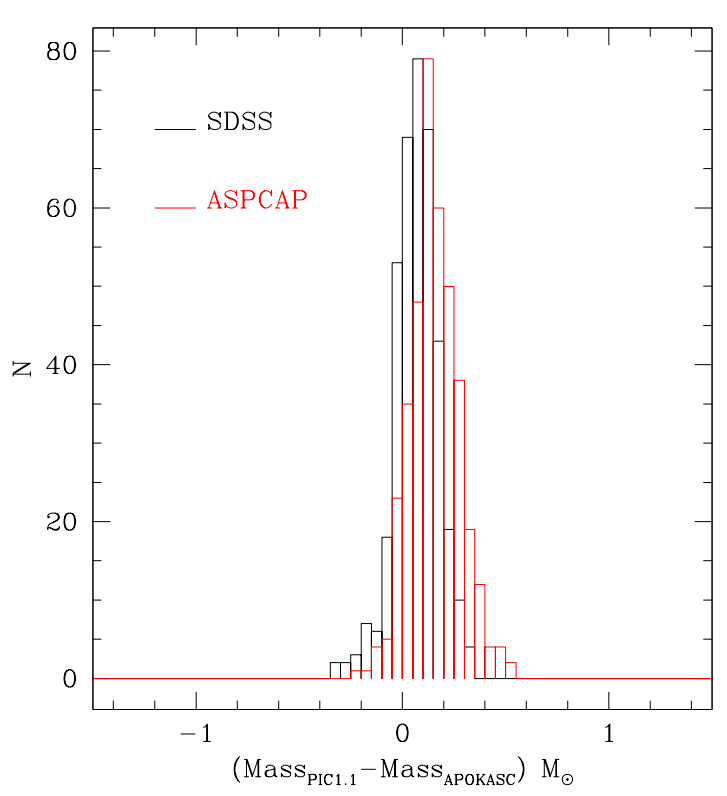}
    \includegraphics[width=0.49\textwidth]{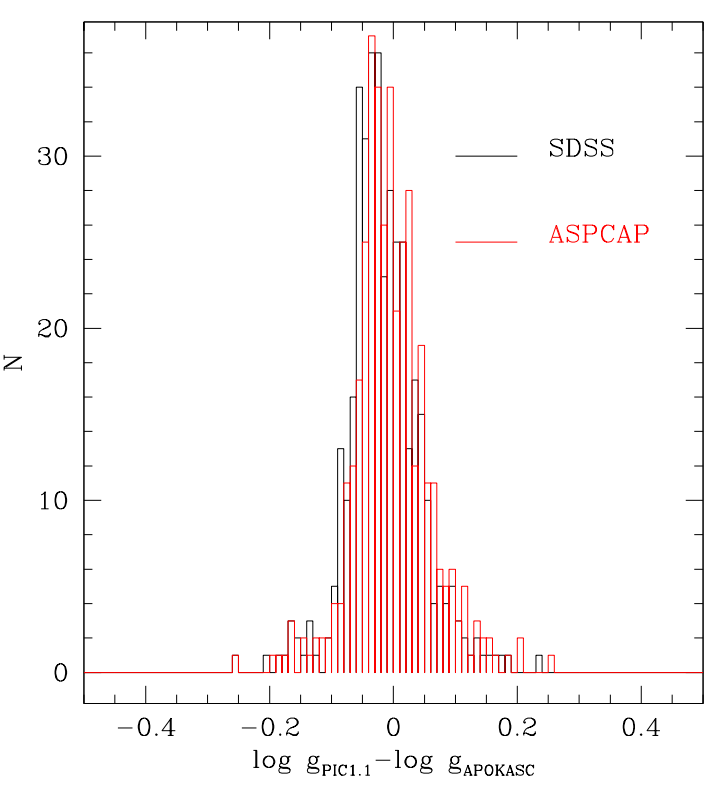}
        \caption{Distributions of effective temperatures, stellar radii, masses and gravities differences between asPIC1.1 and APOKASC sample and two different temperature scales (SDSS, black and ASPCAP, red) .}
        \label{fig:comparison_APOKASC}
\end{figure*}

\section{Comparisons}
\label{sec:comparisons}

 We compared stellar parameters and reddening in asPIC1.1 with the values reported in the 
TIC \citep{stassun2019}. We used the Candidate Target List v8.01 (CTLv8.01) and cross-matched it with asPIC1.1 using the \emph{Gaia} source ID reported in the CTL catalogue. We considered only those stars for which all stellar parameters and the reddening are defined in both catalogues. 
We identified 2 022 913 stars in common between the two catalogues. 
The median differences and the standard deviations of the differences between asPIC1.1 and TIC temperatures, radii, masses, and reddenings are: $\rm\Delta T_{eff}=(-100\pm300)$~K, $\rm\Delta R_{\star}=(0.05\pm0.07)\, R_{\odot}$, $\rm\Delta M_{\star}=(0.05\pm0.17)\,M_{\odot}$, and $\rm\Delta E(B-V)=(-0.002\pm0.055)$. We note that the distribution of stellar masses appears markedly asymmetric with asPIC having a tendency to overestimate the mass with respect to the TIC. This fact may arise from the different choices adopted for the calibration of stellar masses in the two catalogues. In our case, we adopted the calibration from \cite{moya2018} which accounts both for the temperature and absolute luminosity of the stars, while \cite{stassun2018} and \cite{stassun2019} adopt a calibration based only on the temperatures. The differences between the asPIC1.1 and TICv8 estimated parameters are shown in the histograms in Fig.~\ref{fig:comparison_TIC}.

We then considered the Galactic Archaeology with HERMES (GALAH) survey second data release which contains 342~682 spectroscopically observed stars \citep{buder2018}. We cross-matched GALAH with asPIC1.1 using the {\it Gaia} source ID reported in the catalogue. We found 64~061 matched sources. We compared the effective temperatures and the log g in GALAH with the asPIC1.1 values. In asPIC1.1 we report stellar masses (M) and radii (R). The log g is estimated using log g = log M - 2 log R + 4.4374 \citep{smalley2005}. The median differences between the asPIC1.1 and the GALAH effective temperatures and gravities are:  $\rm\Delta T_{eff}=(4\pm100)$ K and
$\rm\Delta log\,g=(0.1\pm0.2)$ dex, as shown in Fig.~\ref{fig:comparison_GALAHDR2}

We also considered the sample of 1111 FGK dwarf stars from the HARPS GTO program, homogeneously analysed in \cite{adibekyan2012}. We matched the catalogue with asPIC1.1 using angular distances (accounting for proper motions) and found 1102 common sources.
The median differences between the asPIC1.1 and the HARPS effective temperatures and gravities are:  $\rm\Delta T_{eff}=(-70\pm90)$ K and
$\rm\Delta log\,g=(-0.03\pm0.15)$ dex (Fig.~\ref{fig:comparison_HARPS}).

Finally, we compared our parameters with 
the ones reported in the first APOKASC catalogue of spectroscopic and asteroseismic data for dwarfs and subgiants \citep{serenelli2017}.
The cross-match with asPIC1.1 yielded 385 common stars for which all parameters were defined in both catalogues.
The analysis was performed both for the SDSS temperature scale and for the spectroscopic
temperature scale adopted in the APOGEE Stellar Parameters and Chemical Abundances pipeline (ASPCAP). 
In the first case, the median differences and the standard deviations of the differences between asPIC1.1 and APOKASC temperatures, radii, masses, and gravities are: $\rm\Delta T_{eff}=(-130\pm90)$ K, $\rm\Delta R_{\star}=(0.1\pm0.2)\, R_{\odot}$, $\rm\Delta M_{\star}=(0.1\pm0.1)\,M_{\odot}$, and $\rm\Delta log\,g=(-0.02\pm0.06)$, while in the second case: $\rm\Delta T_{eff}=(90\pm130)$ K, $\rm\Delta R_{\star}=(0.1\pm0.1)\, R_{\odot}$, $\rm\Delta M_{\star}=(0.1\pm0.1)\,M_{\odot}$, and $\rm\Delta log\,g=(-0.01\pm0.06)$ (see also Fig.~\ref{fig:comparison_APOKASC}). 

Table~\ref{tab:differences} summarises the results of the comparisons between asPIC1.1 and the stellar catalogues considered in this section and we conclude that the average difference between our effective temperatures and those of the other catalogues is -40 K, the average radius difference is 0.08 R$\rm_{\odot}$, and the average mass difference is 0.08 M$\rm_{\odot}$. Considering the internal errors we estimated in Sect.~\ref{sec:stellar_parameters}, and adding them in quadrature to the estimated external errors coming from the comparisons from other catalogues (see above), we conclude that our overall (internal+external) uncertainties on the stellar parameters determination is 230 K (4$\%$) for the effective temperatures, 0.13 R$\rm_{\odot}$ (9$\%$) for the stellar radii, and 0.13 M$\rm_{\odot}$ (11$\%$) for the stellar masses.

\section{Special target list}
\label{sec:special_list}

With the asPIC~1.1 catalogue we also release a special list of objects that consists of all currently known confirmed planet hosts included in the asPIC1.1. This list has been constructed using the Exo-MerCat tool \citep{alei2020}, which accesses four main public catalogues of exoplanets, and obtain an updated list of the known planets along with their physical and orbital parameters. In particular, this tool considers the Extrasolar Planets Encyclopaedia\footnote{\url{http://exoplanet.eu/}}, the NASA Exoplanet Archive\footnote{\url{https://exoplanetarchive.ipac.caltech.edu/}}, the Exoplanets Orbit Database\footnote{\url{http://exoplanets.org/}}, and the Open Exoplanet catalogue\footnote{\url{http://www.openexoplanetcatalogue.com/}}, each one employing its own cataloguing system. For this reason, before obtaining a uniform merging of the objects of interest, Exo-MerCat performs a proper discrimination among aliases and a standardisation of the different entries provided by the catalogues.
The list of the known exoplanets is matched with the asPIC1.1 catalogue.

The special targets list is a living catalogue: it will be continuously updated at any new release of asPIC, because the number of newly discovered planets will constantly increase over the years, not only before but also during the PLATO mission lifetime.

\section{Conclusions}
\label{sec:conclusions}

In this paper, we present the asPIC1.1 catalogue, a public all-sky catalogue of dwarf and subgiant stars of interest for the PLATO survey, based on the \emph{Gaia} DR2 data release. The asPIC catalogue will be fundamental to identifying the best fields for the PLATO space mission and the most promising targets, analysing the instrumental performances as well as planning and optimising ground-based follow-up studies. This catalogue also represents a valuable resource for the construction of stellar samples optimised for the search of transiting planets.

The catalogue includes a total of 2~675~539 stars among which 2~378~177 are FGK dwarfs and subgiants and 297 362 are M-dwarfs. The median distance of FGK stars in our sample is 428 pc and that for M dwarfs is 146 pc. 

We also show that our selection criteria do not bias the statistical distribution of metallicities, and we analyse the impact that metallicity has on the derivation of stellar parameters.
We derived the reddening of our targets and developed an algorithm to infer stellar fundamental parameters from astrometric and photometric measurements. We show that the overall (internal+external) uncertainties on the stellar parameters determined by our analysis are $\sim$230 K (4$\%$) for the effective temperatures, $\sim$0.1 R$\rm_{\odot}$ (9$\%$) for the stellar radii, and $\sim$0.1 M$\rm_{\odot}$ (11$\%$) for the stellar masses. We  also release a special target list containing all known planet hosts cross-matched with our catalogue.

\section*{Acknowledgements}

The authors are grateful to the referee, Keivan Stassun, for reading the manuscript and providing constructive comments and suggestions.
This work presents results from the European Space Agency (ESA) space mission PLATO. The PLATO payload, the PLATO Ground Segment and PLATO data processing are joint developments of ESA and the PLATO Mission Consortium (PMC). Funding for the PMC is provided at national levels, in particular by countries participating in the PLATO Multilateral Agreement (Austria, Belgium, Czech Republic, Denmark, France, Germany, Italy, Netherlands, Portugal, Spain, Sweden, Switzerland, Norway, and United Kingdom) and institutions from Brazil. Members of the PLATO Consortium can be found at \url{https://platomission.com/}. The ESA PLATO mission website is \url{https://www.cosmos.esa.int/plato}. We thank the teams working for PLATO for all their work. MM, GP, VN, VG, LP, SD, SO, SB, RC, LM, IP acknowledge support from PLATO ASI-INAF agreements n.2015-019-R0-2015 and n. 2015-019-R.1-2018. 
We would like to acknowledge the financial support of INAF (Istituto Nazionale di Astrofisica), Osservatorio Astronomico di Roma, ASI (Agenzia Spaziale Italiana) under contract to INAF: ASI 2014-049-R.0 dedicated to SSDC. This work has made use of data from the European Space Agency (ESA) mission Gaia (\url{https://www.cosmos.esa.int/gaia}), processed by the Gaia Data Processing and Analysis Consortium (DPAC, \url{https://www.cosmos.esa.int/web/gaia/dpac/consortium}). 
Part of this work has been carried out within the framework of the National Centre of Competence in Research PlanetS supported by the Swiss National Science Foundation. E.A. acknowledges the financial support of the SNSF.
CA acknowledges support from the KU\,Leuven Research Council (grant C16/18/005: PARADISE) and from  the BELgian federal Science Policy Office (BELSPO) through PRODEX grants Gaia and PLATO.
JMMH is funded by Spanish State Research Agency grants PID2019-107061GB-C61 and MDM-2017-0737 (Unidad de Excelencia  Mar\'{\i}a de Maeztu CAB).
This work has made use of data from the European Space Agency (ESA) mission
{\it Gaia} (\url{https://www.cosmos.esa.int/gaia}), processed by the {\it Gaia}
Data Processing and Analysis Consortium (DPAC,
\url{https://www.cosmos.esa.int/web/gaia/dpac/consortium}). Funding for the DPAC
has been provided by national institutions, in particular the institutions
participating in the {\it Gaia} Multilateral Agreement.
This paper includes data that has been provided by AAO Data Central (datacentral.aao.gov.au).
The GALAH survey is based on observations made at the Australian Astronomical Observatory, under programmes A/2013B/13, A/2014A/25, A/2015A/19, A/2017A/18. We acknowledge the traditional owners of the land on which the AAT stands, the Gamilaraay people, and pay our respects to elders past and present.
The full catalogue of known planets/candidates retrieved by Exo-MerCat is registered as a Virtual Observatory resource and it is available on TOPCAT \citep{taylor2005}. It can be also provided and customized by a dedicated User Interface, available from a public GitHub repository\footnote{\url{https://gitlab.com/eleonoraalei/exo-mercat-gui}.}.Some of the results in this paper have been derived using the HEALPix \citep{gorski2005} package.


%
\bibliographystyle{aa} 
\bibliography{bibexample} 
%

%
%

\appendix 

\section{Johnson and Gaia DR2 colour transformation}
\label{sec:gaiaV}

\begin{figure}
        \centering
        \includegraphics[width=9cm]{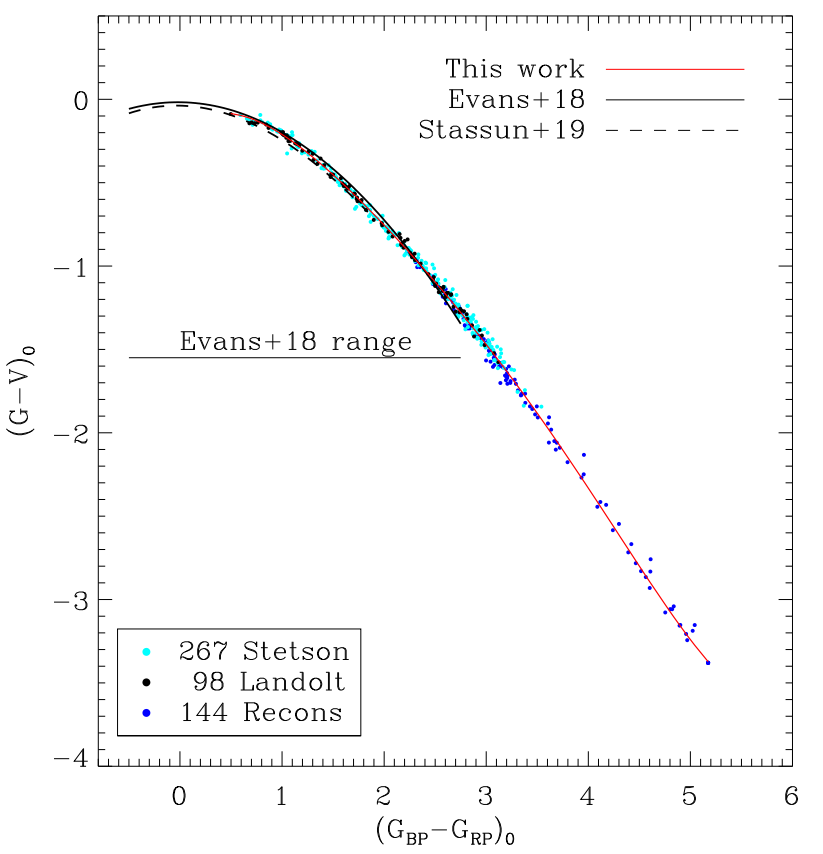}
        \caption{Relationship between the intrinsic ({\it G-V})$\rm_0$ and
        {\it (G$\rm_{BP}$-G$\rm_{RP}$)}$\rm_0$
        colours calibrated in this work (solid red line). We also show the corresponding \citet{evans2018} (solid black line) and \citet{stassun2019} (dashed black line) relationships.}
        \label{fig:gvbprp}
\end{figure}

The definition of PLATO samples requires knowledge of the visual apparent magnitude $V$.
The \textit{V} magnitude we used to select the targets comes from a transformation of \textit{Gaia} DR2 magnitudes and colours to the
Johnson \textit{V} band.
We decided to adopt {\it Gaia} DR2 colours and magnitudes as the basis of our calculation of the {\it V} band magnitude because {\it Gaia} offers
a homogeneous, precise, deep, multi-band, all sky photometry.
\citet{evans2018} give colour--colour relations between several commonly used passbands and \textit{Gaia} DR2 colours. However, the colour range
within which these relationships are applicable does not include all the spectral types of interest for PLATO, in particular the
M-dwarfs. In order to extend the \citet{evans2018} relationship to the entire M-type regime, we used a sample of main sequence stars covering a broad range of spectral types
for which accurate and precise photometry in the Johnson $V$ is given. To this aim, we used photometry of standard stars given in the catalogues of \citet{stetson2000} and \citet{landolt2009}. In addition, for M-dwarfs we  used the RECONS catalogue
 \citep{henry2018} which provides an accurate census of a volume-limited sample of neighbour stars and is a valuable resource in particular for M dwarfs which are poorly represented in the other catalogues. The Stetson, Landolt, and RECONS catalogues were cross-matched with {\it Gaia} DR2. We then corrected each star for reddening using the procedure described in Sect.~\ref{sec:reddening}, and isolated main sequence stars within 200 pc from the Sun with error on the {\it Gaia} {\it G}-band photometry $\sigma_{G}<0.003$, reddening {\it E(B-V)}$<$0.05, and satisfying the astrometric and photometric quality criteria defined by  Eq.~(1) and Eq.~(2) of \citet{arenou2018}. In total we selected 267 stars from  \citet{stetson2000}, 98 stars from \citet{landolt2009} and 144 stars from \citet{henry2018}. We interpolated the  ({\it G-V})$_\textrm{\textrm{0}}$ versus
({\it G$\rm_{BP}$-G$\rm_{RP}$})$\rm_0$
colours with a fifth-order polynomial relation
and the best-fit coefficients are reported in Table~\ref{tab:coeffpolgvbp} of the main text. Figure~\ref{fig:gvbprp} shows a graphical representation of the relationship, overlaying  the best-fit model (solid red line) and the stellar samples that have been used (coloured dots), the \citet{evans2018} relation (solid black line), and the \citet{stassun2019} calibration (dashed black line) used in the TICv8, and based on PHOENIX model atmospheres (provided by K. Stassun, private communication). The standard deviation of the residuals from our best-fit model is $\sigma\rm_{fit}$=0.02745 mag. The relation is valid within 0.5$<$({\it G$\rm_{BP}$-G$\rm_{RP}$}$)\rm_0<$5. We note that this relation is applied to the intrinsic (i.e., reddening-free) magnitudes, and then absorption is added (see Sect.~\ref{sec:reddening}) to obtain the apparent magnitude.

\section{Selection of FGK dwarfs and subgiants using Johnson \textit{B},\textit{V} photometry.}
\label{sec:PIC1.0.0}

\begin{figure}
        \centering
        \includegraphics[width=9cm]{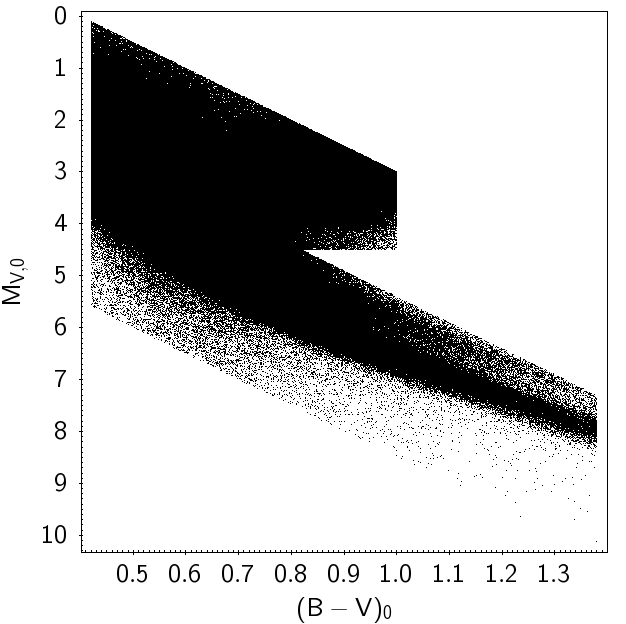}
        \caption{Selection of FGK dwarf and subgiant stars in the Johnson $B$, $V$ absolute intrinsic colour--magnitude diagram.}
        \label{fig:asPIC1.0_selection}
\end{figure}

For completeness of our discussion we report here an alternative selection of FGK dwarf and subgiant stars used during the  construction of the first version of the catalogue (named asPIC1.0). The procedure is similar to the one described in Sect.~\ref{sec:selection_criteria}, but is based on the Johnson photometric system. The selection criteria are then defined in the absolute and intrinsic $M_{V,0}$ versus $(B-V)_0$ colour magnitude diagram and are given by the following inequations where we further distinguish dwarfs from subgiants:
\\
\\
Dwarfs:
\begin{equation*}
   \begin{cases}
    0.42 < (B-V)_0 < 1.38 \\ 
    M_{V,0}\geq 5\,(B-V)_0+0.4 \\ 
    M_{V,0}< 5\,(B-V)_0+3.5 \\
    V<13
    \end{cases}
\end{equation*}
Subgiants:
\begin{equation*}
   \begin{cases}
   0.42 < (B-V)_0 \leq 0.8 \\ 
   M_{V,0}< 5\,(B-V)_0+0.4 \\
   M_{V,0} > 5\,(B-V)_0-2] \\
   V<13
   \end{cases}
\end{equation*}
OR
\begin{equation*}
    \begin{cases}
    0.8 < (B-V)_0 \leq1.0 \\
    M_{V,0}<4.5 \\ 
    M_{V,0}> 5\,(B-V)_0-2 \\
    V<13.
    \end{cases}
\end{equation*}
\noindent
The selection is depicted in Fig.~\ref{fig:asPIC1.0_selection}. One of the limitations of this selection is that 
it cannot be easily extended in the M-dwarf domain because of saturation effects of the $(B-V)$ colour. With the release of {\it Gaia} DR2 photometry we decided to work directly in the {\it Gaia} photometric system, as described in Sect.~\ref{sec:selection_criteria}. Nevertheless the relationships reported  here can be useful to isolate FGK dwarfs and subgiants using Johnson $B$ and $V$ photometry.
The TPR and FPR of this selection are respectively 100$\%$ and 45$\%$, and therefore such a selection has the same effectiveness as that presented in Sect.~\ref{sec:analytic_approximation} in recovering true positives, but implies a larger rate of false positives.

\begin{table*}
\centering
\footnotesize
\caption{
Description of the columns of the asPIC1.1 catalogue.
}
\begin{tabular}{llcl}
\hline
Col. number & Col. name & Units & Description \\
\hline
1 & PICidDR1 & - & PLATO identifier for asPIC1.1 \\
2 & PICnameDR1 & - & PLATO identifier for asPIC1.1 with version number \\
3 & sourceId & - & {\it Gaia} DR2 sourceId \\
4 & RAdeg & deg & Barycentric Right Ascension ($\alpha$) of the source in ICRS at the reference epoch \\
5 & eRAdeg & mas  & Standard error $\rm \sigma_{\alpha*}=\sigma_{\alpha}cos\delta$ of the right ascension in ICRS at the reference epoch\\
6 & DEdeg & deg & Barycentric Declination ($\delta$) of the source in ICRS at the reference epoch\\
7 & eDEdeg & mas & Standard error $\rm \sigma_{\delta}$ of the declination in ICRS at the reference epoch \\
8 & Plx & mas & Parallax \\
9 & ePlx & mas & Standard error on parallax \\
10 & pmRA & mas yr$^{-1}$ & Proper motion in Right Ascension direction 
$\mu_{\alpha*}=\mu_{\alpha}cos\delta$ in ICRS at the reference epoch \\
11 & epmRA & mas yr$^{-1}$ & Standard error on proper motion in right ascension direction \\
12 & pmDE & mas yr$^{-1}$ & Proper motion in Declination direction $\rm\mu_{\delta}$ in ICRS at the reference epoch \\
13 & epmDE & mas yr$^{-1}$ & Standard error on proper motion in declination direction \\
14 & PM & mas yr$^{-1}$ & Total proper motion calculated as PM=$\rm\sqrt{\rm pmRA^2+\rm pmDE^2}$\\
15 & ePM & mas yr$^{-1}$ & Error on the total proper motion \\
16 & Epoch & yr & Reference epoch to which the astrometric source parameters are referred \\
17 & Gflux & e$^{-}$ s$^{-1}$ & Mean flux in the G band \\
18 & eGflux & e$^{-}$ s$^{-1}$ & Error on G band mean flux\\
19 & Gmag & mag & Mean magnitude in the G band \\
20 & eGmag & mag & Error on mean magnitude in the G band\\
21 & BPflux & e$^{-}$ s$^{-1}$  & Mean flux in the integrated BP band\\
22 & eBPflux & e$^{-}$ s$^{-1}$ & Error on mean flux in the integrated BP band\\
23 & BPmag & mag & Mean magnitude in the integrated BP band\\
24 & eBPmag & mag & Error in the mean magnitude in the integrated BP band\\
25 & RPflux & e$^{-}$ s$^{-1}$ & Mean flux in the integrated RP band \\
26 & eRPflux & e$^{-}$ s$^{-1}$ & Error on mean flux in the integrated BP band\\
27 & RPmag & mag & Mean magnitude in the integrated RP band \\
28 & eRPmag & mag & Error in the mean magnitude in the integrated RP band \\
29 & BPRPexcess & - & BP/RP excess factor \\
30 & BPRP & mag & $\rm G_{BP}-G_{RP}$ colour \\
31 & BPG & mag & $\rm G_{BP}-G$ colour \\
32 & GRP & mag & $\rm G-G_{RP}$ colour \\
33 & GLON & deg & Galactic longitude of the object at the reference epoch \\
34 & GLAT & deg & Galactic latitude of the object at the reference epoch \\
35 & ELON & deg & Ecliptic longitude of the object at the reference epoch \\
36 & ELAT & deg & Ecliptic latitude of the object at the reference epoch \\
37 & rest & pc & The estimated distance \citep{bailer2018} \\
38 & rlo & pc & Lower bound of the confidence interval of the estimated distance \citep{bailer2018} \\
39 & rhi & pc & Higher bound of the confidence interval of the estimated distance \citep{bailer2018} \\
40 & erest & pc & Error on the estimated distance calculated as restError$=\rm0.5(rhi-rlo)$ \\
41 & AG & mag & Estimate of the extinction A$\rm_G$ in the G band \\
42 & eAG & mag & Uncertainty on A$\rm_G$ estimate \\
43 & EBPRP & mag & Estimate of reddening E(G$\rm_{BP}$ - G$\rm_{RP}$)\\
44 & eEBPRP & mag &  Uncertainty on reddening E(G$\rm_{BP}$ - G$\rm_{RP}$) \\
45 & EBV & mag & Estimate of reddening E(B-V)\\
46 & eEBV & mag & Uncertainty on E(B-V) reddening\\
47 & extStatus & - & extStatus value. If extStatus=1 the star is outside the reddening map of
\cite{lallement2018}\\
48 & BPRP0 & mag & Dereddened (G$\rm_{BP}$ - G$\rm_{RP}$) colour\\
49 & eBPRP0 & mag & Error on dereddened (G$\rm_{BP}$ - G$\rm_{RP}$) colour (BPRP0)  \\
50 & gaiaV0 & mag & Dereddened apparent visual magnitude {\it V} in the Johnson-Cousin system \\
51 & egaiaV0 & mag & Error on intrinsic visual magnitude (gaiaV0) \\
52 & BJgaiaMV0 & mag & Absolute intrinsic visual magnitude obtained as: BJgaiaMV0=gaiaV0-5$\rm\log_{10}\rm rest+5$\\
53 & eBJgaiaMV0 & mag & Error on absolute intrinsic visual magnitude (BJgaiaMV0) \\
54 & gaiaV & mag & Visual apparent magnitude in the Johnson-Cousin system obtained as: gaiaV=gaiaV0+3.1 EBV\\
55 & egaiaV & mag & Error on apparent visual magnitude {\it V} (gaiaV) \\
56 & BJgaiaMG0 & mag & Absolute intrinsic G magnitude obtained as:  BJgaiaMG0=Gmag-5$\rm\log_{10}\rm rest+5-AG$ \\
57 & eBJgaiaMG0 & mag & Error on intrinsic absolute G magnitude (BJgaiaMG0) \\
58 & Teff & K & Estimate of stellar effective temperature (T$\rm_{eff}$) \\
59 & eTeff & K & Uncertainty on T$\rm_{eff}$ estimate\\
60 & Radius & R$_{\odot}$ & Estimate of stellar radius \\
61 & eRadius & R$_{\odot}$ & Uncertainty on stellar radius estimate \\
62 & Mass & M$_{\odot}$ & Estimate of stellar mass \\
63 & eMass & M$_{\odot}$ & Uncertainty on mass estimate \\
64 & sourceFlag & - & Bitmask with the following meaning: \\
  &  &  & sourceFlag=1 - FGK star \\ 
  &  &  & sourceFlag=2 - M star \\
  &  &  & sourceFlag=4 - Planet Host \\
\hline
\end{tabular}
\label{tab:catalogue}
\end{table*}

\addtocounter{table}{-1}
\begin{table*}
\flushleft
\footnotesize
\caption{
- Continued.
}
\begin{tabular}{llcl}
\hline
Col. number & Col. name & Units & Description \\
\hline
65 & contGaiaMag60 & mag & Total G magnitude of the {\it Gaia} contaminants within 60 arcsec \\
66 & contNumber60 & - & Total number of {\it Gaia} contaminants within 60 arcsec \\
67 & contGaiaMag45 & mag & Total G magnitude of the {\it Gaia} contaminants within 45 arcsec  \\
68 & contNumber45 & - & Total number of {\it Gaia} contaminants within 45 arcsec  \\
69 & contGaiaMag30 & mag & Total G magnitude of the {\it Gaia} contaminants within 30 arcsec\\
70 & contNumber30 & - & Total number of {\it Gaia} contaminants within 30 arcsec \\
\hline
\end{tabular}
\label{tab:catalogue}
\end{table*}

\section{The catalogue}
\label{sec:catalogue}

\subsection{Naming convention}
 The asPIC filename (asPIC.n1.n2) contains two dot-separated integer numbers which are used to distinguish the different versions of the catalogue.
 The first number (n1) indicates an asPIC major release and is linked to the \textit{Gaia} data release used to build the catalogue. When a new \textit{Gaia} release triggers an asPIC major update, n1 is increased by one. The second number (n2) indicates a substantial update, which is not major and thus does not require changing n1. Examples of substantial updates are, for example, a change in (a) the number of objects, (b) the number of columns, (c) the format of one or more columns, or (d) the content of one or more columns. When the first number n1 changes, n2 is set to zero.

\subsection{Content}

In Table~\ref{tab:catalogue} we report the asPIC1.1 column names and a brief description of their meaning. 

\section{Implementation procedure}
\label{sec:implementation}
In this Appendix we describe the implementation procedure adopted to build the asPIC1.1 catalogue.
The implementation of PLATO asPIC1.1 is done using the PLATO Data Processing Centre - ASI (PDPC-A)
infrastructure dedicated to the {\it Gaia} and PLATO missions. 

\begin{itemize}
\item The procedure starts using the Gaia DR2 catalogue main table (gaiaSource) and the Gaia estimated distances table from \cite{bailer2018}. A asPIC main table (asPIC1.1) is created and all sources from \texttt{gaiaSource} where:
\texttt{photGMeanMag} $\leq$ 15.5 AND \texttt{bpRp} is not null AND \texttt{rest} is not null are added.
The asPIC1.1 table includes 52 557 711 sources.

\item The asPIC1.1 table is updated and the following columns are added: \texttt{pmtotal}, \texttt{pmtotalError}, 
\texttt{photGmeanMagError}, \texttt{photBpMeanMagError}, \texttt{photRpMeanMagError}, \texttt{restError}.

\item We created an input file to be fed into the stellar parameters' code (Sect.~\ref{sec:stellar_parameters}). The input file includes all sources in the asPIC1.1 table. The code is run in parallel. The results of the stellar parameters' code (47 966 591 sources) are then fed back into the database in a dedicated table. The stellar parameters' code output columns are: \texttt{ag}, \texttt{agError}, 
\texttt{ebprp}, \texttt{ebprpError}, \texttt{ebv}, \texttt{ebvError}, \texttt{extStatus}, \texttt{teff}, 
\texttt{teffError}, \texttt{radius}, \texttt{radiusError}, \texttt{mass}, \texttt{massError} and \texttt{dwsgFlag}. The flag 
\texttt{dwsgFlag} discriminates  between FGK (\texttt{dwsgFlag}=1), M (\texttt{dwsgFlag}=2) stars and stars that are neither FGK or M (\texttt{dwsgFlag}=0). The main table (asPIC1.1) is updated to include these results.

\item The main table is further updated to include \texttt{bpRp0}, \texttt{bpRp0Error}, \texttt{BJgaiaMG0}, 
\texttt{BJgaiaMG0Error}, \texttt{PICidDR1}, \texttt{gaiaV0}, \texttt{gaiaV0Error}, \texttt{gaiaV}, \texttt{gaiaVError} and \texttt{gaiaMV0}, \texttt{gaiaMV0Error}. Only 47 288 759 sources out of 47 966 591 are updated, since a few sources are left out because they do not fall within the Gaia-Johnson calibration limits. 

\item The definition of the samples in the asPIC110 table is updated according to the final selection criteria:

\begin{enumerate}
\item FGK: (\texttt{dwsgFlag}=1 AND \texttt{gaiaV}<=13.0 AND \texttt{BpRp0}>0.56)
which outputs 2 378 177 sources;
\item M: (\texttt{dwsgFlag}=2 AND \texttt{gaiaV}<=16.0 AND (\texttt{BJgaiaMG0}~>~2.334*(\texttt{BpRp0}) + 2.259)
AND \texttt{rest}<600.0)
which outputs 297~362 sources.
\end{enumerate}
\item Photometric contaminants are searched for and counted in the {\it Gaia} DR2 catalogue for all the candidate targets listed in asPIC using a cone search with 30, 45, and 60 arcsec radius respectively.
To compute the total G magnitude we summed the entire flux of each contaminant. As the contaminants may fall close to the circle border, the computed total magnitude is over-estimated. 
This information was used to compute the columns \texttt{contGaiaMag30}, \texttt{contNumber30}, \newline \texttt{contGaiaMag45},  \texttt{contNumber45},\texttt{contGaiaMag60} and \texttt{contNumber60}.
\end{itemize}

\section{The 3D reddening map}
\label{sec:3d_reddening}

\noindent
The 3D reddening map
presented in \citet{lallement2018} is produced using a hierarchical approach merging individual photometric colour excess measurements of stars with colour excesses estimated from diffuse interstellar bands absorption. These measurements are coupled with a non-homogeneous, large-scale prior reddening distribution  deduced  from  massive  photometric  surveys. 
The 3D map from \citet{lallement2018} was retrieved
from the dedicated website\footnote{http://stilism.obspm.fr}.
The map
is extended in
a region
between $-$2\,kpc and 2\,kpc from the Sun position in the Heliocentric Galactic $X$ and $Y$ coordinates and between $-$300\,pc and 300\,pc in the Galactic
$Z$ coordinate. It has a resolution of $5\,\textrm{pc}^{3}$ and units are $\textrm{mag} \cdot \textrm{pc}^{-1}$.
The map is particularly useful at small distances since it reveals how local
cavities and cloud complexes are spatially distributed around the Sun.

\noindent
We used
the map of \citet{lallement2018} in conjunction with \citet{bailer2018} distances from {\it Gaia} DR2 to estimate the amount of interstellar reddening to our target stars. We evaluated the reddening $E(B-V)_i$ on a sample of $N$, equally spaced, nodal points along each target
direction
by considering the weighted average of the 27 surrounding voxels reddening (a $3 \times 3 \times 3$ voxels cube around the nodal point). The reddening was weighted by the distance of the nodal point from each one of the neighbouring voxels' centers while the distance between two adjacent nodal points was 2\,pc. 
We therefore summed up the reddening along the
boresight
to derive the integrated reddening to the target star

\begin{equation}
    E(B-V)=\sum_{i=1}^{i=N}\, \frac{dE(B-V)_i}{dR} \times \Delta R\,(\textrm{pc}),
\end{equation}

\noindent
where

\begin{equation*}
    \Delta R = 2\,\textrm{pc}
\end{equation*}

\noindent
and

\begin{equation*}
    \frac{dE(B-V)_i}{dR} 
\end{equation*}

\noindent
is the reddening per unit length\footnote{The map reports precisely this quantity.} in the i-th voxel.
The median reddening of asPIC1.1 stars inside the reddening map is equal to $E(B-V)$=0.04 and the median uncertainty is $\sigma\,E(B-V)$=0.02. For the M sample $99.8\%$ of the stars are contained in the reddening map and, for the FGK sample, $81.8\%$. For the stars falling outside the map, reddening is calculated up to the edge of the map, and then a correction is added as described in Sect.~\ref{sec:extension_reddening_map}.

\subsection{Reddening map extension}
\label{sec:extension_reddening_map}

As discussed in the previous section, a small fraction of asPIC
stars falls outside the limits of the \citet{lallement2018} map. For these objects we applied a reddening correction $\Delta E(B-V)$ based on the following model

\begin{equation}
\Delta\,E(B-V)(l,b,d,d_0)=E(B-V)_\infty(l,b)\,\frac{\int_{d_0}^{d}\rho(r)dr}{\int_0^\infty\rho(r)dr},
\end{equation}

\noindent
where $l$ and $b$ are the Galactic longitude and latitude, and the integral is calculated along the ({\it l, b}) direction from the distance \textit{d$_0$} at the border of the map and the distance \textit{d} at which a given star is found. The function $\rho$ is a model of the density distribution of the dust. As in \cite{binney2014}, we used

\begin{equation}
\rho(\textbf{x})=\exp\left[\frac{R_0-R}{h_R}-\frac{|z-z_w|}{k_{fl}h_z}\right],
\end{equation}

\noindent
where \textbf{x} is the position vector of a star with respect to the Galactic centre, \textit{R} the galactocentric distance along the Galactic plane, \textit{z} the orthogonal
distance from the Galactic plane, and k$_{fl}(R)$, z$_w(R)$ describe, respectively, the flaring and warping of the gas disk,

\begin{equation}
k_{fl}(R)=1+\gamma_{fl}\textrm{min}(R_{fl},R-R_{fl}),
\end{equation}

\begin{equation}
z_{w}(R,\phi)=\gamma_{w}\textrm{min}(R_{w},R-R_{w})\sin\phi,
\end{equation}

\noindent
where $\phi$ is the Galactocentric azimuth that increases in the direction of Galactic rotation with the Sun at $\phi$= 0. Table~\ref{tab:parameters_dust_galactic_model} reports the parameter values.
The reddening at infinity $E(B-V)_\infty$ in any given direction is taken from the Schlegel et al.~(1998) map, multiplied by the correction factor reported in \citet{binney2014}:

\begin{table}
	\centering
    \caption{Parameters of the analytical dust's density distribution. Distances are in kiloparsecs.}	
	\begin{tabular}{c c c c c c}
	\hline
    $h_r$ & $h_z$ & $R_{fl}$ & $\gamma_{fl}$ & $R_w$ & $\gamma_w$ \\
    \hline
    4.2 & 0.088 & 1.12$R_0$ & 0.0054 & 8.4 & 0.18 \\
    \hline
    \end{tabular}
    \label{tab:parameters_dust_galactic_model}
\end{table}

\begin{equation}
f(E(B-V)_\infty)=0.6+0.2\left[1-\tanh\left(\frac{E(B-V)_\infty-0.15}{0.3}\right)\right].
\end{equation}

\noindent
The correction described in this section is applied only if the reddening calculated up to the \citet{lallement2018} map edge was not larger than the reddening E(B-V)$_\infty$ estimated from the \citet{schlegel1998} map. 

\section{The impact of reddening and absorption on the determination of the stellar radius}
\label{sec:impact_of_reddening_on_stellar_radius}

\begin{figure}
        \centering
        \includegraphics[width=0.45\textwidth]{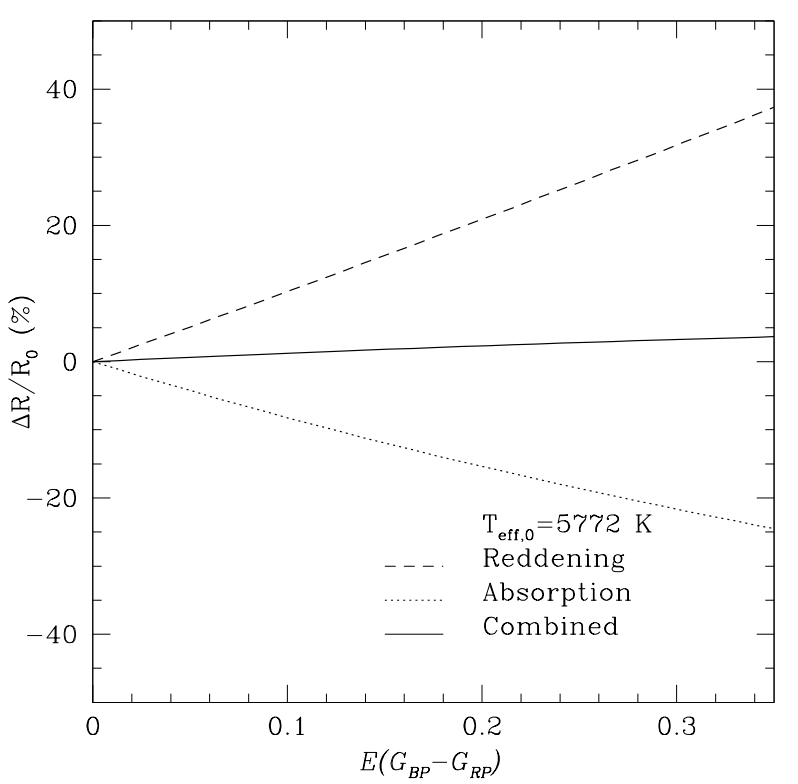}
        \caption{
        Percentage variation of the apparent radius $R$ estimated neglecting the presence both of absorption and reddening, with respect to the true radius $R_0$ of a solar-type star (with effective temperature T$\rm_{eff}$=5772 K) as a function of reddening $E(G_\textrm{BP}-G_\textrm{RP})$ (solid line). The dashed and dotted lines represent the apparent radius variation resulting from neglecting either the reddening (dashed) or the absorption (dotted), respectively.}
        \label{fig:apparent_radius}
\end{figure}

It is of fundamental importance to estimate the impact of reddening and absorption on the determination of the stellar radius, a crucial parameter in the context of transiting planet searches. \\
By applying the Stefan-Boltzmann law it follows that the relative variation of the apparent stellar radius $R$, estimated neglecting the presence of dust, with respect to the true stellar radius $R_0$ can be expressed as

\begin{equation}
\label{eq:apparent_radius}
    \frac{\Delta R}{R_0}=\left ( \frac{T_{\textrm{eff},0}}{T_{\textrm{eff}}} \right )^2\,10^{-0.2(A + \Delta BC)} - 1,
\end{equation}

\noindent
where $T_{\textrm{eff},0}$ is the true effective temperature, and $T_\textrm{eff}$ the estimated temperature of the target star, while $A$ and $\Delta\,BC$ are the absorption and differential bolometric correction in a given photometric band, respectively. \\
Considering a reddening law, a colour-effective temperature relation, and an effective temperature-bolometric correction relation, it is possible to express the relative radius variation as a function of reddening only.
For example, adopting the relations
discussed in Section~\ref{sec:stellar_parameters}, valid for the \textit{Gaia} photometric bands, we obtain that, in correspondence to a reddening equal to $E(G_\textrm{BP}-G_\textrm{RP})=0.3$ mag, the radius of a solar-type star would be overestimated by about $3\%$ if both reddening and absorption were neglected in the radius calculation, as shown in Fig.~\ref{fig:apparent_radius} (solid line). The individual contributions resulting from neglecting either the reddening or the absorption are illustrated respectively by the dashed and dotted lines in the same figure. Therefore, in correspondence to the same amount of reddening,  $E(G_\textrm{BP}-G_\textrm{RP})=0.3$ mag, neglecting the effect of the reddening leads to underestimation of the effective temperature which in turn leads to overestimation of the stellar radius by about 32$\%$ (dashed line). On the other hand, neglecting the effect of absorption produces an underestimation of the stellar luminosity, which implies an underestimation of the stellar radius by about 22$\%$ (dotted line).  
{ Therefore, reddening and absorption have two opposite effects on the apparent stellar radius which are partially compensating each other.} 

\begin{figure}
        \centering
        \includegraphics[width=0.45\textwidth]{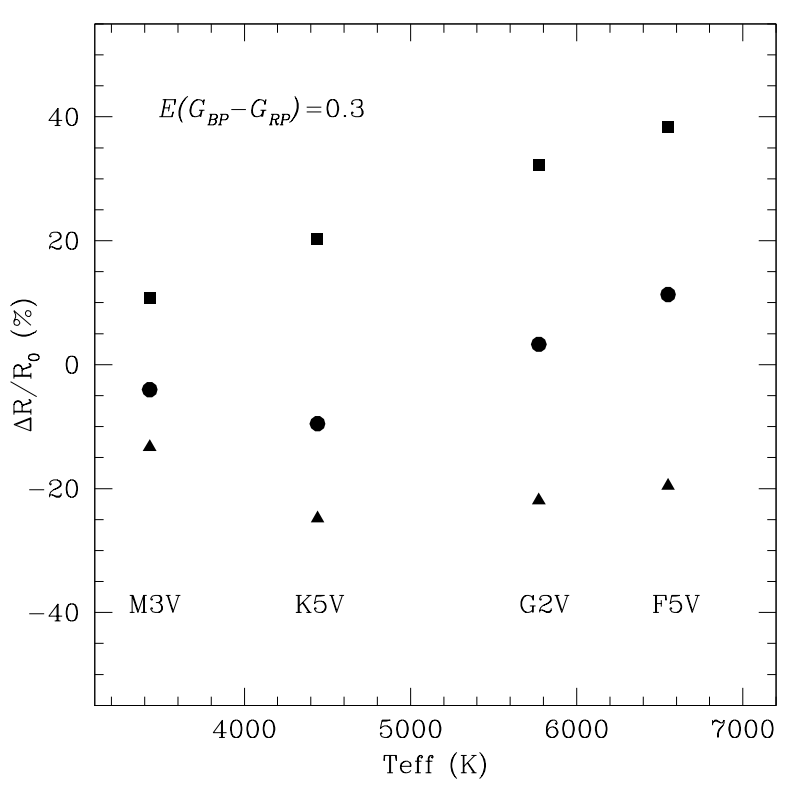}
        \caption{Percentage variation of the apparent radius
        $R$ estimated neglecting the presence of  both absorption and reddening (dots) for stars of different spectral types and for a reddening equal to {\it E(G$_{BP}$-G$_{RP}$)}=0.3. Squares and triangles represent the percentage radius variation resulting from neglecting the reddening or the absorption, respectively.}
        \label{fig:apparent_radius_sptype}
\end{figure}

These results hold for a solar-type star with effective temperature $T\rm_{eff}$=5772 K. It should be kept in mind that this compensation effect depends on the spectral type of the target source, the slope of the adopted colour--effective temperature relation (and therefore the adopted photometric system), and the reddening law. For example, in Fig.~\ref{fig:apparent_radius_sptype} we show
that
for an M3V star, neglecting absorption would produce an underestimation of the stellar radius by about 13$\%$, neglecting the reddening would produce an overestimation of the stellar radius by about
11$\%$, while neglecting both of them would underestimate the radius by about 4$\%$. For a F5V star,  neglecting absorption would produce an underestimation of the stellar radius by about 20$\%$, neglecting the reddening would produce an overestimation of the stellar radius by about
38$\%$, while neglecting both of them would overestimate the radius by about 11$\%$. 

The presence of interstellar matter has a smaller impact on M-dwarfs given that these sources are intrinsically redder than earlier type stars, and absorption and reddening are more prominent at shorter wavelengths. Moreover, as we move towards F-type stars, the effect of reddening becomes dominant over the effect of absorption (because of the strong dependence on effective temperature in Eq.~\ref{eq:apparent_radius} and the steepening of the colour--effective temperature relation, Sect.~\ref{sec:intrinsic_colour_effective_temperature}) 
which generally leads to an overestimation of the stellar radius.

\end{document}